\documentclass[a4paper,11pt]{article}
\usepackage[left=2.2cm,right=2.2cm,top=2cm,bottom=2.5cm]{geometry}
\renewcommand{\arraystretch}{1.2} 
\usepackage{caption}
\usepackage[utf8]{inputenc}
\usepackage{multirow}
\usepackage[pdftex]{graphicx}
\usepackage{amssymb}
\usepackage{amsmath}
\usepackage{cite}
\usepackage{comment}
\usepackage{braket}
\usepackage{slashed}
\usepackage{longtable}
\usepackage{booktabs}
\usepackage{array}
\usepackage[normalem]{ulem}
\usepackage{subcaption}
\usepackage{needspace}

\usepackage[dvipsnames]{xcolor}
\usepackage{xspace}

\usepackage[]{hyperref}
\usepackage{cleveref}
\allowdisplaybreaks
\numberwithin{equation}{section}

\newcommand{\appsection}[1]{%
  \refstepcounter{APP}%
  \section*{Appendix \theAPP: #1}%
  \addcontentsline{toc}{section}{\theAPP~#1}%
  \setcounter{equation}{0}%
}

\newcommand{\refapp}[1]{\hyperref[app:#1]{Appendix~\ref*{app:#1}}}
\newcommand{\reffig}[1]{\hyperref[fig:#1]{Fig.~\ref*{fig:#1}}}
\newcommand{\refeq}[1]{\hyperref[eq:#1]{Eq.~(\ref*{eq:#1})}}
\newcommand{\refeqs}[2]{\hyperref[eq:#1]{Eqs.~(\ref*{eq:#1})-(\ref*{eq:#2})}}
\newcommand{\refeqa}[2]{\hyperref[eq:#1]{Eqs.~(\ref*{eq:#1})} and~\hyperref[eq:#2]{(\ref*{eq:#2})}}
\newcommand{\refsec}[1]{\hyperref[sec:#1]{Section~\ref*{sec:#1}}}
\newcommand{\refsubsec}[1]{\hyperref[sec:#1]{Subsection~\ref*{sec:#1}}}
\newcommand{\reftab}[1]{\hyperref[tab:#1]{Table~\ref*{tab:#1}}}

\newcommand{\C}{\mathcal{C}}

\newcommand{\K}{\mathcal{K}}

\newcommand{\X}{\mathcal{X}}
\newcommand{\Y}{\mathcal{Y}}
\newcommand{\W}{\mathcal{W}}

\newcommand{\Tr}{\text{Tr}}

\newcommand{\sdot}{\!\cdot\!}

\newcounter{TODO}

\begin{document}

\thispagestyle{empty}

\begin{flushright}
{\small
CERN-TH-2026-144\\
SI-HEP-2026-12\\
P3H-26-044\\
MITP-26-029\\
KA-TP-12-2026\\
September 22, 2026 
}
\end{flushright}

\vspace*{-2mm}
\begin{center}
\fontsize{15}{20}\selectfont
\bf
\boldmath
Three-particle di-light-cone distribution amplitudes\\
of the $B$-meson in heavy-quark effective theory
    
\end{center}

\vspace{2mm}

\begin{center}

{Riccardo Bartocci$^{a,b}$, 
Philipp B\"oer$^c$, 
Thorsten Feldmann$^d$, \\
Max Ferr\'e$^e$, 
Nico Gubernari$^f$, 
Daniel Vladimirov$^d$ }\\[5mm]
{\it\small

{$^{\, a}$}Institute for Theoretical Physics, Karlsruhe Institute of Technology, \\Wolfgang-Gaede-Str. 1, 76131 Karlsruhe,
Germany
\\[2mm]
{$^{\, b}$}Institute for Astroparticle Physics, Karlsruhe Institute of Technology, \\76344 Eggenstein-Leopoldshafen,
Germany
\\[2mm]
{$^{\, c}$}CERN, Theoretical Physics Department, CH-1211 Geneva 23, Switzerland
\\[2mm]
{$^{\, d}$}Theoretische Physik 1, Center for Particle Physics Siegen, \\
Universit\"at Siegen, 57068 Siegen, Germany
\\[2mm]
{$^{\, e}$}PRISMA$^{++}$ Cluster of Excellence \& Mainz Institute for Theoretical Physics\\
Johannes Gutenberg University, Staudingerweg 9, D-55128 Mainz, Germany
\\[2mm]
{$^{\, f}$}Helmholtz-Institut f\"ur Strahlen- und Kernphysik, Universit\"at Bonn, 53115 Bonn, Germany
\\[4mm]

E-mail:{\textnormal{ \texttt{%
    \,riccardo.bartocci@kit.edu, 
    \,philipp.boeer@cern.ch, 
    \,thorsten.feldmann@uni-siegen.de, 
    \,ferremax@uni-mainz.de, 
    \,nicogubernari@gmail.com, 
    \,daniel.vladimirov@uni-siegen.de
}}}
}
\end{center}

\vspace{1mm}
\begin{abstract}\noindent

\noindent
We present a systematic study of the three-particle di-light-cone distribution amplitudes (DLCDAs) of the $B$-meson.
They are defined through $B$-meson--to--vacuum matrix elements of trilocal HQET operators, in which the light antiquark and the gluon field-strength tensor are located on two back-to-back light rays.
In this sense, the DLCDAs generalise the conventional $B$-meson light-cone distribution amplitudes to configurations where soft fields couple to collinear degrees of freedom in two distinct directions.
As such, they parametrise the non-perturbative dynamics associated with non-factorisable soft-gluon contributions in rare and non-leptonic exclusive $B$-meson decays.
We derive the complete Lorentz decomposition of the matrix elements of generic trilocal operators, identify eight independent DLCDAs, and organise them in a basis of definite twist. 
Using local operator identities and equations-of-motion constraints, we obtain tree-level relations for their normalisation integrals and first moments in terms of a minimal set of hadronic parameters. 
These relations allow us to construct simple momentum-space models for all independent DLCDAs. 
For the leading-twist distribution, we further incorporate the perturbative radiative tail at order $\alpha_s$ and discuss its impact on the resulting parametrisation.
\end{abstract}

\newpage

\tableofcontents

\section{Introduction}
\label{sec:Introduction}

Exclusive decays of $b$ hadrons provide precision probes of the flavour sector of the Standard Model and are among the most promising avenues for discovering new particles and interactions. The presence of various widely separated energy scales, $m_W \gg m_b \gg \Lambda_{\rm QCD}$, makes their theoretical description complicated. After integrating out the electroweak-scale degrees of freedom ($m_W \to \infty$) one is left with hadronic matrix elements of operators in the effective electroweak Hamiltonian. A quantitative understanding of the QCD dynamics of those matrix elements poses the main theoretical challenge. 
An expansion in inverse powers of the heavy $b$-quark mass often allows one to express
them in terms of a minimal set of simpler non-perturbative quantities as well as perturbatively calculable short-distance coefficients.

For decays into light and energetic particles, where the recoil energy is of the same order as the $b$-quark mass, the separation of short- and long-distance scales is achieved in the heavy-mass limit within the QCD factorisation approach 
(QCDF)~\cite{Beneke:1999br,Beneke:2000ry,Beneke:2001ev}. Alternatively, light-cone sum rules (LCSRs) employ light-cone operator product expansions to relate correlation functions to simpler hadronic quantities in certain kinematic limits; see, for example, the reviews in Refs.~\cite{Colangelo:2000dp,Khodjamirian:2023wol}. In both approaches, light-cone distribution amplitudes (LCDAs) provide the essential hadronic input. For heavy hadrons, these LCDAs describe light-like correlations between the light quarks, gluons and the static heavy quark, and parametrise the soft long-distance dynamics relevant for the corresponding exclusive decays at large recoil.

The simplest and best-known example in this context is the leading-twist two-particle $B$-meson LCDA $\phi_B^+(\omega;\mu)$, which has been studied extensively over the past decades; see, e.g., Refs.~\cite{Grozin:1996pq,Beneke:2000wa,Kawamura:2001jm,Descotes-Genon:2002crx,Lange:2003ff,Braun:2003wx,Khodjamirian:2005ea,Lee:2005gza,Beneke:2011nf,Bell:2013tfa,Feldmann:2014ika,Braun:2019wyx,Beneke:2021rjf,Wang:2021yrr,Feldmann:2022uok}.
Its higher-twist counterparts have been classified systematically in Ref.~\cite{Braun:2017liq} (see also Refs.~\cite{Braun:2015pha,Descotes-Genon:2009jif}) including the corresponding three-particle LCDAs with an additional gluon field-strength tensor aligned on the same light-cone. Whereas in QCDF theorems, typically only the first inverse moment $\lambda_B$ of $\phi_B^+(\omega;\mu)$ appears at leading order, also the detailed shape of the LCDAs becomes relevant when radiative corrections are included, or in phenomenological applications using LCSRs.

The organisation of power corrections in terms of the conventional higher-twist $B$-meson LCDAs is justified as long as soft partons in the short-distance process couple only to energetic particles that propagate (close to) a single light cone. This picture changes in processes that involve more than one energetic direction. Whereas the leading interaction of soft gluons with energetic partons is eikonal, and colour transparency guarantees the decoupling from colour-neutral particles, this argument does not generally hold beyond the eikonal limit. In this case, the standard single-ray non-local operators that define the multi-particle $B$-meson LCDAs need to be generalised to soft functions involving fields along multiple light cones. In this work we restrict ourselves to the simplest case, involving two back-to-back light-like directions, and we refer to the corresponding $B$-meson matrix elements as di-light-cone distribution amplitudes (DLCDAs).

Objects of this type recently appeared in several contexts. First, in the study of QED corrections to two-body exclusive $B$-decays in Refs.~\cite{Beneke:2019slt,Beneke:2020vnb,Beneke:2022msp,Cornella:2026zkd}, where the sensitivity to the second light-cone direction arises from the non-decoupling of eikonal photons from electrically charged external particles. In QCD, the leading three-particle DLCDA has been introduced in Ref.~\cite{Qin:2022rlk} to describe charm-loop contributions in rare $B_{d,s} \to \gamma \gamma$ decays, where the charm-quark mass acts as an intermediate hard-collinear scale. 
This function would also be needed 
for estimating non-factorisable soft-gluon contributions to $\bar{B}_s^0 \to D_s^+ \pi^-$ and $\bar{B}^0 \to D^+ K^-$ decays, 
which are considered to be among the theoretically cleanest non-leptonic decay channels of the $B$-meson (for a simplified LCSR approach, see the recent analysis in Ref.~\cite{Piscopo:2023opf}).
Three-particle DLCDAs for the $\Lambda_b$ baryon
enter the calculation of non-factorisable contributions of 
four-quark operators to rare $\Lambda_b \to \Lambda \ell^+\ell^- $ transitions, which again has been shown in the context of LCSRs 
\cite{Feldmann:2023plv}.
Lastly, even 
four- and five-particle DLCDAs arise in the factorisation of weak-annihilation amplitudes~\cite{Boeer:CKM2025,BoeerNeubertStillger}.

At present, relatively little is known about the properties of the DLCDAs. The analysis of the one-loop evolution kernel of the QED-generalised $B$-meson LCDA in Ref.~\cite{Beneke:2022msp} revealed various new features that indicate that the DLCDAs are a qualitatively different class of functions from the conventional LCDAs. 
The derivation of the one-loop anomalous dimension of the leading three-particle DLCDA has confirmed that these features also arise in QCD~\cite{Huang:2023jdu}.
For example, the coupling of soft gluons to the two directions leads to rescattering phases, such that the DLCDAs become complex-valued functions. Another surprising new feature is that the analytic structure of the DLCDAs also allows for negative light-cone momenta of light quarks and gluons. 
However, a recent study of the one-loop renormalisation-group equation (RGE)~\cite{Bartocci:2024bbf} has shown that, when combined with the analytic properties of the relevant hard-scattering kernels in a given factorisation theorem, it is possible in practical applications to restrict the integration domain to positive light-cone momenta.

In this work, we study the simplest DLCDAs of the $B$-meson in QCD, defined by an operator containing a heavy-quark field, a light-antiquark field on the light cone $n^\mu$, and a gluon field-strength tensor on the opposite light cone $\bar{n}^\mu$. We provide a complete classification of the corresponding three-particle DLCDAs, generalising the work of Ref.~\cite{Geyer:2005fb}, and organise them in a basis of definite twist. We then derive constraints from the equations of motion for local matrix elements, which translate at tree level into relations among the normalisation integrals and first moments of the DLCDAs. On this basis, we construct simple models for all independent distributions that can be used for first phenomenological studies. For the leading-twist DLCDA, we further show how the tree-level parametrisation is modified by the perturbative radiative tail at first order in the strong coupling, adapting the strategy recently developed in Ref.~\cite{Feldmann:2025dcs}.

The remainder of this paper is organised as follows. In \refsec{defDDAs} we define the relevant trilocal matrix element, derive the complete basis of three-particle DLCDAs, and relate it to a twist decomposition. In \refsec{models} we derive the local constraints and present a class of simple phenomenological models. The treatment of the leading-twist radiative tail is discussed in \refsec{radiative_tail}. We conclude in \refsec{concl}.

\section{Definition of the three-particle DLCDAs}
\label{sec:defDDAs}

In this section, we introduce for the first time the complete set of three-particle $B$-meson DLCDAs that parametrise the trilocal hadronic matrix element
\begin{align}
    \label{eq:TME}
    \langle 0| \bar q(z_1 n)\, gG_{\mu\nu}(z_2 \bar{n})\Gamma h_v(0) |\bar B(v)\rangle \,.
\end{align}
Throughout this work, all non-local fields are understood to implicitly include the corresponding gauge-link factors:
$
    \bar q(z_1 n) \mapsto \bar q(z_1 n)\,[z_1 n,0] \,,
$
where $[z_1 n,0]$ denotes a straight Wilson line along the $n^\mu$ light cone. An analogous prescription applies to the gluon field, with $G_{\mu\nu}(z_2 \bar n) \mapsto [0,z_2\bar{n}]
\, G_{\mu\nu}(z_2 \bar n) \, [z_2\bar{n},0]$.
In addition, $\Gamma$ denotes a Dirac matrix and $h_v$ is the effective heavy-quark field with velocity $v_\mu$.
The vectors $n$ and $\bar{n}$ are light-cone vectors satisfying $n^2=\bar{n}^2=0$, which we choose as 
\begin{align}
    v_\mu = \frac{1}{2}\left(n_\mu + \bar n_\mu\right), \qquad
    n \cdot \bar n = 2\,.
\end{align}
Quantities transverse to both $n^\mu$ and $\bar n^\mu$ are denoted by a subscript~$\perp$, for example $p_\perp^\mu$ and $\gamma_\perp^\mu$.
The parameters $z_1$ and $z_2$ denote the positions of the antiquark and the gluon on the respective light cone.

To determine the number of independent DLCDAs that parametrise the matrix element~\eqref{eq:TME} we follow the procedure outlined in Ref.~\cite{Geyer:2005fb}.
This amounts to constructing the most general Lorentz decomposition of
$\langle 0| \bar q(z_1 n)\, gG_{\mu\nu}(z_2 \bar{n})\gamma_5\gamma_\rho\sigma_{\alpha\beta} h_v(0) |\bar B(v)\rangle$
in terms of independent tensor structures.
In particular, we identify all possible structures compatible with Lorentz invariance for a rank-five tensor featuring two antisymmetric index pairs:
\begin{equation}
\begin{aligned}
    \label{eq:Gennbar}
    &
    \frac{
        \langle 0| \bar q(z_1 n)\, gG_{\mu\nu}(z_2 \bar{n})\gamma_5\gamma_\rho\sigma_{\alpha\beta} h_v(0) |\bar B(v)\rangle
    }{
        \langle 0| \bar q(0)\,  \gamma_5 h_v(0) |\bar B(v)\rangle
    }
    =
    \\* &\hspace*{3cm}
        +g_{\alpha[\mu}\, g_{\nu]\beta}\, g_{\rho\sigma}\, \K^\sigma_1
        +
        g_{\sigma[\mu}\, g_{\nu][\alpha}\, g_{\beta]\rho}\, \K^\sigma_2
        +
        g_{\sigma[\alpha}\, g_{\beta][\mu}\, g_{\nu]\rho}\, \K^\sigma_3
    \\* &\hspace*{3cm}
        + g_{\kappa[\mu}\,g_{\nu]\tau}\,
        g_{\lambda[\alpha}\,{g_{\beta]}}^\tau\,g_{\rho\sigma}
        \left(
            n^\kappa n^\lambda\,\K^\sigma_4
            +
            \bar{n}^\kappa \bar{n}^\lambda\,\K^\sigma_5
        \right)
    \\* &\hspace*{3cm}
        +
        \bar{n}_{[\mu } n_{\nu]} \,
        g_{\rho[\alpha}\,g_{\beta]\sigma}\,\K^\sigma_6
        + 
        \bar{n}_{[\alpha}  n_{\beta]} \,
        g_{\rho[\mu}\, g_{\nu]\sigma}\,\K^\sigma_7
    \\* &\hspace*{3cm}
        + g_{\kappa[\mu}\,g_{\nu]\tau}\,g_{\lambda[\alpha}\,{g_{\beta]}}^\tau\,g_{\rho\sigma}
        \bigg(
            \bar{n}^\kappa n^\lambda\,\K^\sigma_8
            +n^\kappa \bar{n}^\lambda\,\K^\sigma_9
        \bigg)
        + 
        \bar{n}_{[\mu } n_{\nu]}\,\bar{n}_{[\alpha} n_{\beta]} \,
        g_{\rho\sigma}\, \K^\sigma_{10}
    \,.
\end{aligned}
\end{equation}
Here $\K^\sigma_i := n^\sigma\, \widetilde{\K}_i +  \bar{n}^\sigma\, \widehat{\K}_i$ and each DLCDA is understood to be a function of $z_1$ and $z_2$, e.g., \hbox{$\widetilde{\K}_1 := \widetilde{\K}_1(z_1,z_2)$}.
We also abbreviate $A_{[\mu}B_{\nu]}=A_\mu B_\nu-A_\nu B_\mu$.
For convenience, we adopt the following convention for the phase and the normalisation of states, although most of our results are independent of these choices:
\begin{align}
    \langle 0| \bar q(0)\,  \gamma_5 h_v(0) |\bar B(v)\rangle
    =
    - i\, m_B F_B(\mu) 
    \,,
\end{align}
where $F_B(\mu)$ is the  heavy-quark effective theory (HQET) decay constant, which is related to the physical $B$-meson decay constant (see, e.g., Ref.~\cite{Braun:2017liq}).\footnote{
    Note that this definition differs from that of Ref.~\cite{Braun:2017liq} by a factor of $m_B$. One factor of $\sqrt{m_B}$ originates from the convention for the state normalisation, while the other arises from the definition of the HQET decay constant:
    $
    \langle 0| \bar q(0)\,  \gamma_5 h_v(0) |\hat{\bar{B}}(v)\rangle
    =
    - i F_B^{\rm BJM17} (\mu) \, .
    $
}
The twenty DLCDAs appearing in \refeq{Gennbar} are not independent; rather, they are subject to constraints arising from the equations of motion (EOM) and from identities among the Dirac matrices, such as the Chisholm identity (see Ref.~\cite{Geyer:2005fb} and \refapp{genDAs}).
These constraints imply that the following DLCDAs must vanish
\begin{equation}
\begin{aligned}
\widetilde{\K}_4 &= 0,&
\widehat{\K}_5 &= 0,&
\widetilde{\K}_8 &= 0,&
\widehat{\K}_9 &= 0,&
\widetilde{\K}_{10} &= 0,&
\widehat{\K}_{10} &= 0.
\end{aligned}
\end{equation}
In addition, the following relations between DLCDAs hold:
\begin{equation}
\begin{aligned}
\widetilde{\K}_1 &= \widetilde{\K}_3,&
\widehat{\K}_1 &= \widehat{\K}_3,&
\widehat{\K}_4 &= -\widetilde{\K}_7 = -\widetilde{\K}_9,&
 \widetilde{\K}_5 &= \widehat{\K}_7 = - \widehat{\K}_8 .
\end{aligned}
\end{equation}
The four DLCDAs on the left-hand side of these identities can be considered independent.
The following four DLCDAs, which are unconstrained and hence independent, complete our basis of eight DLCDAs:
\begin{equation}
\begin{aligned}
&
\widetilde{\K}_{2},\qquad
\widehat{\K}_{2},\qquad
\widetilde{\K}_{6},\qquad
\widehat{\K}_{6}.
\end{aligned}
\end{equation}

While the procedure of Ref.~\cite{Geyer:2005fb} is useful for determining the number of independent DLCDAs, it does not directly provide a parametrisation of the matrix element in \refeq{TME}, which is required for practical applications.
To this end, we introduce the following parametrisation of \refeq{TME}, inspired by Eq.~(2.7) of Ref.~\cite{Braun:2017liq}:
\begin{align}
\begin{aligned}
    \label{eq:Brnnbar}
    \langle 0| \bar q(z_1 n) & gG_{\mu\nu}(z_2 \bar{n})\Gamma h_v(0) |\bar B(v)\rangle =
    \\*
    &+
    \frac12 m_B F_B(\mu)  \Tr\biggl\{\gamma_5 \Gamma 
    P_+
    \biggl[
    \left(n_\mu \gamma_\nu - n_\nu \gamma_\mu\right) Y_x
    + 
    \left(\bar{n}_\mu \gamma_\nu - \bar{n}_\nu \gamma_\mu\right) Y_y
    \\*
    &
    + i\,\epsilon_{\mu\nu\alpha\beta}\,n^\alpha \gamma^\beta \gamma_5\,\widetilde{Y}_x 
    + i\,\epsilon_{\mu\nu\alpha\beta}\,\bar{n}^\alpha \gamma^\beta \gamma_5\,\widetilde{Y}_y 
    \\*
    & 
    + (n_\mu \gamma_\nu-n_\nu \gamma_\mu)\slashed{n}\,{Z_x} 
    + (\bar{n}_\mu \gamma_\nu-\bar{n}_\nu \gamma_\mu)\slashed{\bar{n}}\,{Z_y} 
    \\*
    & 
    - i\,\epsilon_{\mu\nu\alpha\beta}\,n^\alpha \bar{n}^\beta\gamma_5 \, \widetilde{X}_{xy}
    - \left(n_\mu \bar{n}_\nu-n_\nu \bar{n}_\mu\right)\slashed{n}\, W_{xy}
    \biggr]\biggr\}(z_1, z_2;\mu)\,,
\end{aligned}
\end{align}
where $P_+ := \frac{1+\slashed v}{2}$ is the projector onto the heavy-quark spinor.
Throughout, we adopt the following conventions
\begin{align}
    g_{\mu\nu}
    = {\rm diag} (+,-,-,-)
    \,,\quad
    \sigma_{\mu\nu}
    =
    \frac{i}{2}
    [\gamma_\mu,\gamma_\nu]
    \,,\qquad
    \gamma_5 = i\,\gamma^0 \gamma^1 \gamma^2 \gamma^3 \, ,
    \qquad
    \epsilon^{0123}=-1
    \,,
\end{align}
in agreement with Ref.~\cite{Braun:2017liq}.
To verify that the eight DLCDAs appearing in the equation above are indeed independent, we relate them to those introduced in \refeq{Gennbar}.
This can be achieved by applying the procedure outlined in \refapp{genDAs}.
We find
\begin{equation}
\begin{aligned}
    Y_x 
    &= 
    -2\widehat{\K}_4 + 2\widehat{\K}_6 + \widetilde{\K}_2 - 2\widetilde{\K}_5 \,,
    &\qquad\qquad
    Y_y 
    &= 
    \widehat{\K}_2 - 2\widehat{\K}_6 \,,
    \\
    \widetilde{Y}_x 
    &= 
    2\widehat{\K}_6 - \widetilde{\K}_1 - 2\widetilde{\K}_5 \,,
    &
    \widetilde{Y}_y 
    &= 
    -\widehat{\K}_1 - 2\widehat{\K}_6 + 2\widetilde{\K}_5 \,,
    \\
    Z_x 
    &= 
    -2\bigl(\widehat{\K}_4 - \widehat{\K}_6 + \widetilde{\K}_5\bigr) \,,
    &
    Z_y 
    &= 
    -2\widehat{\K}_6 \,,
    \\
    \widetilde{X}_{xy} 
    &= 
    2\bigl(\widehat{\K}_6 - \widetilde{\K}_5\bigr) \,,
    &
    W_{xy} 
    &= 
    -\widehat{\K}_4 + \widehat{\K}_6 - \widetilde{\K}_5 - \widetilde{\K}_6 \,.
\end{aligned}
\end{equation}
Since this system of equations can be inverted, the basis chosen in \refeq{Brnnbar} is indeed valid.
\\

Following Ref.~\cite{Braun:2017liq}, DLCDAs of definite twist can be defined by means of appropriate projectors, reflecting the familiar power-counting rules from soft-collinear effective theory (SCET).
For the light-quark field, the leading- and subleading-power projections are
\begin{align}
    \frac{\slashed{\bar n}\slashed n}{4}\,q(z_1 n)
    \qquad \text{and} \qquad
    \frac{\slashed n\slashed{\bar n}}{4}\,q(z_1 n)
    \,,
\end{align}
respectively.
For the gluon field strength, it is convenient to distinguish transverse Lorentz indices, denoted by $\mu_\perp,\nu_\perp$, which are orthogonal to both $n^\mu$ and $\bar n^\mu$.
The relevant projections are then
\begin{align}
    \bar n^\mu G_{\mu\nu_\perp}(z_2 \bar n)
    \,, \qquad
    \bar n^\mu n^\nu G_{\mu\nu}(z_2 \bar n)
    \,, \qquad
    G_{\mu_\perp\nu_\perp}(z_2 \bar n)
    \,, \qquad
    n^\mu G_{\mu\nu_\perp}(z_2 \bar n)
    \,.
\end{align}
The first of these is leading power, the next two are subleading, and the last one is sub-subleading.
Starting from these projections, we construct the projectors that define operators of definite twist.
At twist three we find
\begin{equation}
2m_B F_B(\mu) \,\Phi_{3}^{(n\bar n)}(z_1,z_2;\mu)
=
\langle 0|\,
\bar q(z_1 n)\, g G_{\mu\nu}(z_2 \bar n)\,\bar n^\nu\, \slashed{n} \,\gamma_{\perp}^\mu \gamma_5\, h_v(0)
\,|\bar B(v)\rangle ,
\end{equation}
where
\begin{equation}
    \Phi_{3}^{(n\bar n)} =
    2\left(Y_x - \widetilde{Y}_x\right) \,.
\end{equation}
At twist four we find
\begin{align}
2m_B F_B(\mu) \,\Phi_{4}^{(n\bar n)}(z_1,z_2;\mu)
&=
\langle 0|\,
\bar q(z_1 n)\, g G_{\mu\nu}(z_2 \bar n)\,\bar n^\nu\, \slashed{\bar{n}} \,\gamma_{\perp}^\mu \gamma_5\, h_v(0)
\,|\bar B(v)\rangle, 
\\[2mm]
2m_B F_B(\mu) \,\Psi_{4}^{(n\bar n)}(z_1,z_2;\mu)
&=
\langle 0|\,
\bar q(z_1 n)\, g G_{\mu\nu}(z_2 \bar n)\, \bar{n}^\mu n^\nu\, \slashed{n} \gamma_5\, h_v(0)
\,|\bar B(v)\rangle,
\\[2mm]
2m_B F_B(\mu) \,\widetilde\Psi_{4}^{(n\bar n)}(z_1,z_2;\mu)
&=
\langle 0|\,
\bar q(z_1 n)\, i g \widetilde G_{\mu\nu}(z_2 \bar n)\, \bar{n}^\mu n^\nu\, \slashed{n} \, h_v(0)
\,|\bar B(v)\rangle ,
\end{align}
where
\begin{align}
    \widetilde G_{\mu\nu}
    :=
    \frac{1}{2}
    \epsilon_{\mu\nu\alpha\beta}
    G^{\alpha\beta}
\end{align}
and
\begin{align}
    \Phi_{4}^{(n\bar n)} & =
    2 \left(\widetilde{Y}_x + Y_x - 2Z_x\right)
    \,,\\
    \Psi_{4}^{(n\bar n)} & =
    2 Y_y
    \,,\\
    \widetilde\Psi_{4}^{(n\bar n)} & =
    -2 \left(\widetilde{X}_{xy} + \widetilde{Y}_y\right)
    \,.
\end{align}
At twist five we find
\begin{align}
2m_B F_B(\mu) \,\Phi_{5}^{(n\bar n)}(z_1,z_2;\mu)
&=
\langle 0|\,
\bar q(z_1 n)\, g G_{\mu\nu}(z_2 \bar n)\, n^\nu\, \slashed{n} \,\gamma_{\perp}^\mu \gamma_5\, h_v(0)
\,|\bar B(v)\rangle,
\\[2mm]
2m_B F_B(\mu) \,\Psi_{5}^{(n\bar n)}(z_1,z_2;\mu)
&=
\langle 0|\,
\bar q(z_1 n)\, g G_{\mu\nu}(z_2 \bar n)\, \bar{n}^\mu n^\nu\, \slashed{\bar{n}} \gamma_5\, h_v(0)
\,|\bar B(v)\rangle,
\\[2mm]
2m_B F_B(\mu) \,\widetilde\Psi_{5}^{(n\bar n)}(z_1,z_2;\mu)
&=
\langle 0|\,
\bar q(z_1 n)\, i g \widetilde G_{\mu\nu}(z_2 \bar n)\, \bar{n}^\mu n^\nu\, \slashed{\bar{n}} \, h_v(0)
\,|\bar B(v)\rangle ,
\end{align}
where
\begin{align}
    \Phi_{5}^{(n\bar n)} & =
    2 \left(\widetilde{Y}_y + Y_y - 2Z_y\right)
    \,,\\
    \Psi_{5}^{(n\bar n)} & =
    2 \left(2W_{xy} - Y_x\right)
    \,,\\
    \widetilde\Psi_{5}^{(n\bar n)} & =
    -2 \left(\widetilde{X}_{xy} - \widetilde{Y}_x\right)
    \,.
\end{align}
At twist six we find
\begin{equation}
2m_B F_B(\mu) \,\Phi_{6}^{(n\bar n)}(z_1,z_2;\mu)
=
\langle 0|\,
\bar q(z_1 n)\, g G_{\mu\nu}(z_2 \bar n)\, n^\nu\, \slashed{\bar{n}} \,\gamma_{\perp}^\mu \gamma_5\, h_v(0)
\,|\bar B(v)\rangle ,
\end{equation}
where
\begin{equation}
    \Phi_{6}^{(n\bar n)} =
    2\left(Y_y - \widetilde{Y}_y\right) \,.
\end{equation}

\section{Models and constraints}
\label{sec:models}

Practical calculations, such as the evaluation of hadronic amplitudes in factorisation frameworks, require information about \emph{the shape}
of the DLCDAs.
Since these distributions are not known from first principles, phenomenological analyses require models that incorporate the available non-perturbative information while respecting the theoretical constraints.

In this section, we derive the constraints implied by the EOM and use them to obtain tree-level relations among the first moments, thereby reducing the number of independent inputs. We then implement these constraints in a class of exponential models multiplied by a polynomial.

Before proceeding, we summarise the approximations adopted in this section. Radiative corrections induce a tail at large momenta, which spoils a consistent normalisation of the DLCDAs. We therefore neglect these contributions and defer their discussion to \refsec{radiative_tail}. We further omit contributions from four-particle DLCDAs and restrict ourselves to positive support for all distributions, as discussed in the Introduction.

The local limit $z_1,z_2 \to 0$ defines the normalisation of the DLCDAs. 
For the DLCDAs of definite twist introduced in \refsec{defDDAs}, one finds
\begin{align}
F^{(n\bar n)}(0,0)=N_{[F]},
\qquad 
F \in \{\Phi_3,\Phi_4,\Psi_4,\widetilde{\Psi}_4,
\Phi_5,\Psi_5,\widetilde{\Psi}_5,\Phi_6\}\,,
\end{align}
where the corresponding normalisation constants are
\begin{align}
\label{eq:normalisation}
\begin{aligned}
N_{\Phi_3} &= \frac{1}{3}(\lambda_E^2+\lambda_H^2), \qquad\qquad
& N_{\Phi_4} &= \frac{1}{3}(\lambda_E^2-\lambda_H^2), \\
N_{\Psi_4} &= \frac{\lambda_E^2}{3}, \qquad\qquad
& N_{\widetilde{\Psi}_4} &= \frac{\lambda_H^2}{3}, \\
N_{\Phi_5} &= \frac{1}{3}(\lambda_E^2-\lambda_H^2), \qquad\qquad
& N_{\Psi_5} &= -\frac{\lambda_E^2}{3}, \\
N_{\widetilde{\Psi}_5} &= -\frac{\lambda_H^2}{3}, \qquad\qquad
& N_{\Phi_6} &= \frac{1}{3}(\lambda_E^2+\lambda_H^2).
\end{aligned}
\end{align}
The chromoelectric and chromomagnetic hadronic parameters $\lambda_E^2$ and $\lambda_H^2$ are defined through the local quark--gluon matrix element~\cite{Braun:2017liq}:
\begin{align}\label{eq:2deriv_local_antisym}
\langle 0|\bar q\, g G_{\mu\nu}\, \Gamma\, h_v|\bar B(v)\rangle
=
-\frac{1}{6}m_B F_B(\mu) \, \mathrm{Tr}\Big\{
\gamma_5 \Gamma P_+
\Big[
i\,\lambda_H^2\sigma_{\mu\nu}
+
(\lambda_H^2-\lambda_E^2)
(v_\mu\gamma_\nu-v_\nu\gamma_\mu)
\Big]
\Big\}\,.
\end{align}
Numerical estimates for these parameters can be found in Refs.~\cite{Grozin:1996pq,Nishikawa:2011qk,Rahimi:2020zzo}. These matrix elements constitute the antisymmetric part, i.e. $i gG_{\mu \nu}:=[i\overleftarrow{D}_\mu,i\overleftarrow{D}_\nu]$,  of the local matrix elements containing two covariant derivatives acting on the spectator quark. Supplemented by the residual mass of the $B$-meson HQET state $\bar{\Lambda}=m_B-m_b$, those hadronic quantities fully parametrise the symmetric part of such matrix elements~\cite{Grozin:1996pq}
\begin{align}\label{eq:2deriv_local_sym}
\begin{aligned}
\langle 0|\bar q\, \frac12\left\{i\overleftarrow{D}_\mu,i\overleftarrow{D}_\nu\right\}\, \Gamma\, h_v|\bar B(v)\rangle
&=
-\frac{i}{6} m_B F_B(\mu) \, \mathrm{Tr}\bigg\{
\gamma_5 \Gamma P_+
\Big[
(6 \bar{\Lambda}^2+2 \lambda_E^2+\lambda_H^2)v_\mu v_\nu\\
&-(\bar{\Lambda}^2+\lambda_E^2+\lambda_H^2)g_{\mu\nu}-\left(\bar{\Lambda}^2+\frac{\lambda_E^2}{2}\right)(\gamma_\mu v_\nu +\gamma_\nu v_\mu)\Big]\bigg\}\,.
\end{aligned}
\end{align}

In full generality, the local matrix element of the $B$-meson with three covariant derivatives can be decomposed into 14 independent Lorentz structures, which we will use in the following to derive the EOM relations and compute the first moments of the corresponding distribution amplitudes. Explicitly, one has
\begin{align}
\label{eq:three_derivative_parametrisation}
\begin{aligned}
&\langle 0 |\bar q \, i\overleftarrow{D}_\rho \, i \overleftarrow{D}_\nu \, i\overleftarrow{D}_\mu \, \Gamma \,h_v|\bar{B}(v)\rangle= \\
& -\frac{i}{2}m_B F_B(\mu) \, \mathrm{Tr}\bigg\{
\gamma_5 \Gamma P_+\Big[
  \big(d \, v_\mu v_\nu v_\rho + e_1 \, g_{\mu\nu} v_\rho + e_2 \, g_{\mu\rho} v_\nu + e_3 \, g_{\nu\rho} v_\mu \big)  \\
& + \big(f_1 \,  g_{\mu\nu} \gamma_\rho + f_2 \, g_{\mu\rho} \gamma_\nu + f_3 \, g_{\nu\rho} \gamma_\mu \big)  + \big(g_1 \, v_\mu v_\nu \gamma_\rho + g_2 \, v_\mu v_\rho \gamma_\nu + g_3 \, v_\nu v_\rho \gamma_\mu \big) \\
& + \big(h_1 \, i\sigma_{\mu\nu} v_\rho + h_2 \, i\sigma_{\mu\rho} v_\nu + h_3 \, i\sigma_{\nu\rho} v_\mu \big)  +  \, h_4 \, \epsilon_{\mu\nu\rho\sigma} v^\sigma
i \gamma_5\Big]\bigg\}.
\end{aligned}
\end{align}
For a generic three-particle DLCDA $F^{(n\bar n)}(\omega_1,\bar{\omega}_2)$ in position space, we define its Fourier transform as
\begin{equation}
\label{eq:def_fourier}
    F^{(n\bar n)}(z_1,z_2)
    := 
    \int_0^\infty d\omega_1 \int_0^\infty d\bar{\omega}_2 \, 
    e^{-i\omega_1 z_1 -i\bar{\omega}_2 z_2}
    f^{(n\bar n)}(\omega_1,\bar{\omega}_2)\,,
\end{equation}
where $f \in \{\phi_3,\phi_4,\psi_4,\widetilde{\psi}_4,
\phi_5,\psi_5,\widetilde{\psi}_5,\phi_6\}$. 
We then define the moments of $f$ as 
\begin{equation}
\label{eq:def_moments}
\langle \omega_1^m \, \bar{\omega}_2^n \rangle_{[F]} := 
\int_0^\infty d\omega_1 \int_0^\infty d\bar{\omega}_2 \, 
\omega_1^m \bar{\omega}_2^n \, f^{(n\bar n)}(\omega_1,\bar{\omega}_2)\,.
\end{equation}

Based on the parametrisation introduced in~\refeq{three_derivative_parametrisation}, the first moments of the three-particle DLCDAs with respect to the light-quark momentum can be derived from
\begin{align} 
\label{eq:w1_moments}
\langle 0 \big| \bar q \, i \overleftarrow{D}_\rho \, \left[\,i\overleftarrow{D}_\nu,i\overleftarrow{D}_\mu\,\right] \, \Gamma \, h_v \big| \bar{B}(v) \rangle \, n^\rho \,,
\end{align} 
while the first moments with respect to the gluon momentum can be derived from
\begin{align} 
\label{eq:w2_moments}
\langle 0 \big| \bar q \, \left[\left[\,i \overleftarrow{D}_\rho , i \overleftarrow{D}_\nu\,\right],i\overleftarrow{D}_\mu\right] \, \Gamma \, h_v \big| \bar{B}(v) \rangle \, \bar n^\mu \,.
\end{align} 
Explicit expressions for these first moments are collected in \refapp{DA_first_moments}.

Starting from the general parametrisation of the three-derivative local matrix element in \refeq{three_derivative_parametrisation}, one can apply the light-quark and heavy-quark EOM to derive additional constraints among the 14 independent Lorentz structures describing this matrix element.
For the light quark, we have
\begin{equation}
    \langle 0 |\bar q \,i \overleftarrow{\rlap{\hspace{0.75mm}/}{D}}\, i \overleftarrow{D}_\nu\, i\overleftarrow{D}_\mu \, \Gamma \,h_v|\bar{B}(v)\rangle 
     =0\,.
\end{equation}
Contracting this matrix element with appropriate Dirac structures (e.g., $\Gamma=\gamma^\lambda \gamma_5 i\sigma^{\alpha\tau}$) yields a set of five independent relations among the parameters. 
Similarly, the heavy-quark EOM relates the three-derivative matrix element to a two-derivative one:
\begin{equation}\label{eq:hEOM}
    \langle 0 |\bar q\,  i\overleftarrow{D}_\nu\, i\overleftarrow{D}_\rho \,i (v\cdot \overleftarrow{D})\, \Gamma\,h_v|\bar{B}(v)\rangle
     =\bar{\Lambda} \,\langle 0 |\bar q \,i\overleftarrow{D}_\nu\, i\overleftarrow{D}_\rho\, \Gamma \,h_v|\bar{B}(v)\rangle \, .
\end{equation}
The matrix element on the right-hand side is parametrised in terms of the hadronic parameters $\lambda_E^2$, $\lambda_H^2$, and $\bar{\Lambda}$ and it can be obtained from \refeqa{2deriv_local_antisym}{2deriv_local_sym}.
Evaluating \refeq{hEOM} for $\Gamma=\gamma^\lambda\, \gamma_5\,i\,\sigma^{\alpha\tau}$ provides a second set of 5 independent relations among the 14 parameters.
In total, the EOM constraints eliminate 8 of the 14 parameters, expressing them in terms of the remaining ones and the hadronic parameters $\bar{\Lambda}$, $\lambda_E^2$, and $\lambda_H^2$.
Explicit constraints for the parameters are provided in \refapp{DA_first_moments}, where the remaining independent parameters are $e_1, f_3, g_1, g_2, g_3, h_2$. Additional constraints can be derived employing the EOM of the gluon field-strength tensor $G_{\mu\nu}$, in particular we can write
\begin{equation}
    \label{eq:Dadj}
  \left[ \left[ i \overleftarrow{D}_\rho,i \overleftarrow{D}_\nu \right],i \overleftarrow{D}_\mu\right] :=
  g \, {\cal D}_\mu^{\rm adj} G_{\nu\rho} \,.
\end{equation}
Additionally, neglecting four-particle contributions, the relation $i {\cal D}_\mu^{\rm adj} G^{\mu\rho} \approx 0$ holds, providing two further independent relations that allow one to eliminate $f_3$ and $g_3$.
The full set of explicit expressions for all EOM constraints and the resulting reduced parametrisation is provided in \refapp{DA_first_moments}. 

This approach ensures that all first moments of the three-particle DLCDAs are consistently expressed in terms of a minimal set of independent hadronic parameters. Furthermore, the EOM constraints induce relations among the first moments of the DLCDAs, so that all moments can be expressed in terms of just four independent ones. We choose these to be $\langle \omega_1 \rangle_{\Phi_3}$, $\langle \omega_1 \rangle_{\Phi_4}$, $\langle \omega_1 \rangle_{\Psi_4}$ and $\langle \omega_1 \rangle_{\Phi_5}$. These expressions for the first moments in $\omega_1$ read: 
\begin{align}
\label{eq:w1_moment_relations-Psi4}
\langle \omega_1 \rangle_{\widetilde{\Psi}_4} &= -2 \, \langle \omega_1 \rangle_{\Phi_4} + \langle \omega_1 \rangle_{\Psi_4},\\
\langle \omega_1 \rangle_{\Psi_5} &= \frac{1}{2} \Big( - \langle \omega_1 \rangle_{\Phi_3} + \langle \omega_1 \rangle_{\Psi_4} - \langle \omega_1 \rangle_{\Phi_5} \Big), \\
\langle \omega_1 \rangle_{\widetilde{\Psi}_5} &= \frac{1}{6} \Big( - 3 \, \langle \omega_1 \rangle_{\Phi_3} - 6 \, \langle \omega_1 \rangle_{\Phi_4} + 3 \, \langle \omega_1 \rangle_{\Psi_4} + 3 \, \langle \omega_1 \rangle_{\Phi_5} \Big), \\
\langle \omega_1 \rangle_{\Phi_6} &= \langle \omega_1 \rangle_{\Psi_4} - \langle \omega_1 \rangle_{\Phi_4},
\end{align}
while for the moments in $\bar{\omega}_2$ we obtain
\begin{align}
\langle \bar{\omega}_2 \rangle_{\Phi_3} &= \frac{1}{4} \Big( -3 \, \langle \omega_1 \rangle_{\Phi_3} + 3 \, \langle \omega_1 \rangle_{\Phi_4} - 3 \, \langle \omega_1 \rangle_{\Psi_4} + 2 \, \bar{\Lambda} \, \lambda_E^2 + 2 \, \bar{\Lambda} \, \lambda_H^2 \Big), \\
\langle \bar{\omega}_2 \rangle_{\Phi_4} &= \frac{1}{4} \Big( -3 \, \langle \omega_1 \rangle_{\Phi_4} - 3 \, \langle \omega_1 \rangle_{\Phi_5} + 2 \, \bar{\Lambda} (\lambda_E^2 - \lambda_H^2) \Big), \\
\langle \bar{\omega}_2 \rangle_{\Psi_4} &= \frac{1}{12} \Big( -3 \, \langle \omega_1 \rangle_{\Phi_3} - 3 \, \langle \omega_1 \rangle_{\Psi_4} - 3 \, \langle \omega_1 \rangle_{\Phi_5} + 4 \, \bar{\Lambda} \, \lambda_E^2 \Big), \\
\langle \bar{\omega}_2 \rangle_{\widetilde{\Psi}_4} &= \frac{1}{12} \Big( -3 \, \langle \omega_1 \rangle_{\Phi_3} + 6 \, \langle \omega_1 \rangle_{\Phi_4} - 3 \, \langle \omega_1 \rangle_{\Psi_4} + 3 \, \langle \omega_1 \rangle_{\Phi_5} + 4 \, \bar{\Lambda} \, \lambda_H^2 \Big), \\
\langle \bar{\omega}_2 \rangle_{\Phi_5} &= \frac{1}{12} \Big( -3 \, \langle \omega_1 \rangle_{\Phi_4} - 3 \, \langle \omega_1 \rangle_{\Phi_5} + 2 \, \bar{\Lambda} (\lambda_E^2 - \lambda_H^2) \Big), \\
\langle \bar{\omega}_2 \rangle_{\Psi_5} &= \frac{1}{12} \Big( 3 \, \langle \omega_1 \rangle_{\Phi_3} + 3 \, \langle \omega_1 \rangle_{\Psi_4} + 3 \, \langle \omega_1 \rangle_{\Phi_5} - 4 \, \bar{\Lambda} \, \lambda_E^2 \Big), \\
\langle \bar{\omega}_2 \rangle_{\widetilde{\Psi}_5} &= \frac{1}{12} \Big( 3 \, \langle \omega_1 \rangle_{\Phi_3} - 6 \, \langle \omega_1 \rangle_{\Phi_4} + 3 \, \langle \omega_1 \rangle_{\Psi_4} - 3 \, \langle \omega_1 \rangle_{\Phi_5} - 4 \, \bar{\Lambda} \, \lambda_H^2 \Big), \\
\langle \bar{\omega}_2 \rangle_{\Phi_6} &= \frac{1}{12} \Big( -3 \, \langle \omega_1 \rangle_{\Phi_3} + 3 \, \langle \omega_1 \rangle_{\Phi_4} - 3 \, \langle \omega_1 \rangle_{\Psi_4}  + 2 \, \bar{\Lambda} \, \lambda_E^2 + 2 \, \bar{\Lambda} \, \lambda_H^2 \Big).
\label{eq:w2_moment_relations-Phi6}
\end{align}

We now proceed to model the DLCDAs by the factorised ansatz
\begin{equation}
    \label{eq:fmodel}
    f^{(n\bar n)}(\omega_1,\bar{\omega}_2) 
    = N_{[F]} \, \varphi_1^{[f]}(\omega_1) \, \varphi_2^{[f]}(\bar{\omega}_2)\,,
\end{equation}
with 
\begin{align}
\varphi_1^{[f]}(\omega_1) &= \frac{1}{k!} \, \left(\frac{\omega_1}{\omega_0} \right)^k \, \frac{e^{-\frac{\omega_1}{\omega_0}}}{\omega_0} \,
\frac{1+b_{[f]} \, \omega_1/\omega_0}{1+(1+k)\, b_{[f]}} \, \theta(\omega_1)
\,,
\\
\varphi_2^{[f]}(\bar{\omega}_2) &= \frac{1}{\bar k!} \, \left(\frac{\bar{\omega}_2}{\bar \omega_0} \right)^{\bar k} \, \frac{e^{-\frac{\bar{\omega}_2}{\bar\omega_0}}}{\bar\omega_0} \,
\frac{1+c_{[f]} \, \bar{\omega}_2/\bar\omega_0}{1+(1+\bar k)\, c_{[f]}} \, \theta(\bar{\omega}_2)
\,.
\end{align} 
Here, the functional dependence is built as a product of a polynomial prefactor and an exponential suppression at large values of light-cone momenta,
where we allow for two distinct auxiliary reference scales $\omega_0$ and $\bar \omega_0$ for the quark 
and gluon momenta, respectively.\footnote{The need for different reference scales for quarks and gluons in the DLCDAs is also supported by the properties of the renormalisation-group equations, as discussed in the next section.}
Notice that at this stage, the exponential fall-off ensures the existence of all non-negative moments of the DLCDAs, and the Heaviside functions restrict the momenta to positive values.
However, this behaviour will be altered as soon as we take into account the effects associated with the so-called ``radiative tail'', see  \refsec{radiative_tail}.
The powers $k$ and $\bar k$ in the above ansatz
have to be chosen to reproduce the expected scaling of the corresponding 
DLCDA in the limit $\omega_1,\bar{\omega}_2 \to 0$.  This is determined by the 
conformal spin of the constituent fields~\cite{Braun:1989iv}, and the explicit values read\footnote{
    The relation between the conformal spin and behaviour at small $\omega$ values is only meaningful if we restrict ourselves to the positive support~\cite{Bartocci:2024bbf}. 
}
\[
(k,\bar k) =
\begin{cases}
(1,2) & \phi_3,\\[1ex]
(0,2) & \phi_4,\\[1ex]
(1,1) & \psi_4, \widetilde{\psi}_4,\\[1ex]
(1,0) & \phi_5,\\[1ex]
(0,1) & \psi_5, \widetilde{\psi}_5,\\[1ex]
(0,0) & \phi_6.
\end{cases}
\]
Deviations from the minimal exponential model are encoded in the dimensionless shape parameters $b_{[f]}$ and $c_{[f]}$, 
which introduce linear deformations in $\omega_1$ and $\bar{\omega}_2$, respectively. The specific 
normalisation factors in the denominators, $1+(1+k)b_{[f]}$ and $1+(1+\bar k)c_{[f]}$, are chosen such that 
the normalisation of the distribution equals $N_{[F]}$. 
Owing to the factorised structure at the initial scale, the same form is preserved under renormalisation group evolution, as long as one restricts oneself 
to situations where these distributions are inserted into a proper factorisation formula~\cite{Bartocci:2024bbf}. 

\begin{figure}[t]
    \centering
    \begin{subfigure}{0.47\textwidth}
        \centering
        \includegraphics[width=\linewidth]{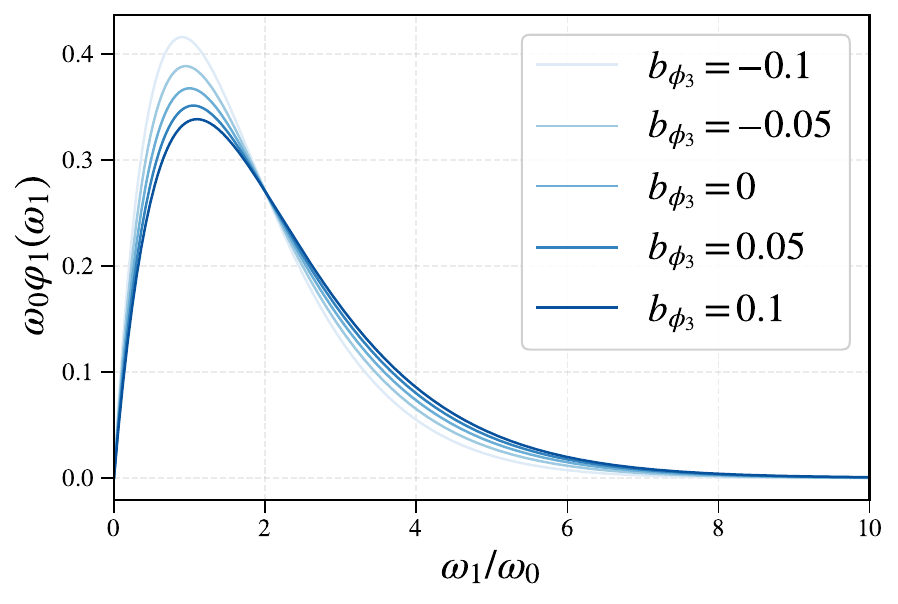}
    \end{subfigure}
    \hfill
    \begin{subfigure}{0.48\textwidth}
        \centering
        \includegraphics[width=\linewidth]{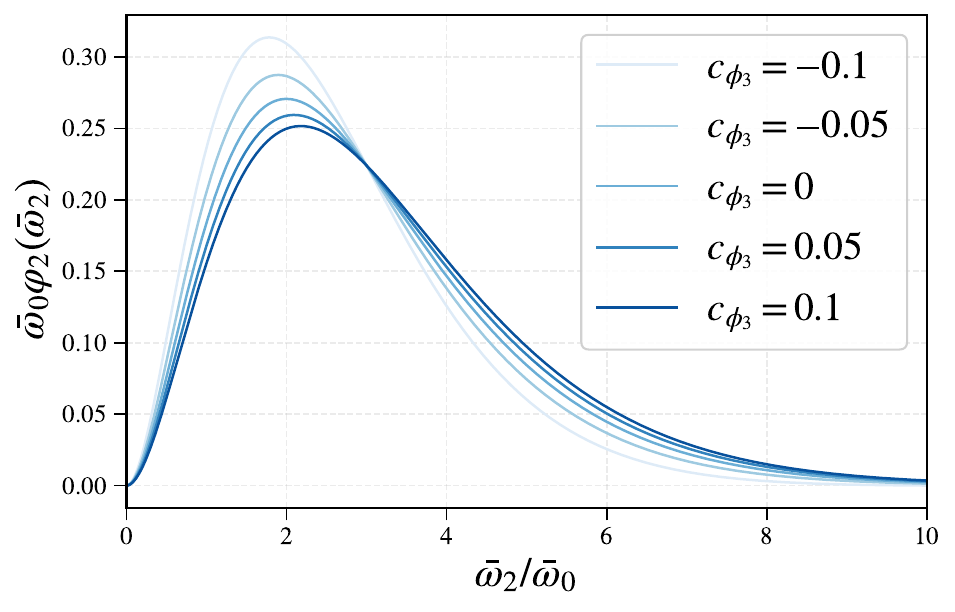}
    \end{subfigure}
    \caption{
   Illustration of the $\omega_1$ and $\bar{\omega}_2$
    dependence of the leading-twist DLCDA $\phi_3$.
    The left panel shows the 
    function $\omega_0\varphi_1(\omega_1)$ 
    as a function of 
    $\omega_1/\omega_0$, for representative values of the parameter 
    $b_{\phi_3}$. 
    The right panel displays the 
    function $\bar{\omega}_0\varphi_2(\bar{\omega}_2)$
    as a function of $\bar{\omega}_2/\bar\omega_0$, for representative values of $c_{\phi_3}$. 
    In both cases, the distributions 
    are rescaled by $\omega_0$ or $\bar\omega_0$, such that only the dependence on the dimensionless 
    ratios $\omega_1/\omega_0$ and $\bar{\omega}_2/\bar\omega_0$ 
    and on the deformation parameters is highlighted.
    }
    \label{fig:phi3_model_plot}
\end{figure}

In \reffig{phi3_model_plot} we show the 
momentum dependence of the DLCDA $\phi_3$, 
as function of the corresponding dimensionless ratios
$\omega_1/\omega_0$ and $\bar{\omega}_2/\bar{\omega}_0$. Here and in the following, we use the shorthand notation $\varphi_{1,2}:=\varphi_{1,2}^{{\phi_3}}$ for brevity.
The left panel shows the function $\omega_0\varphi_1(\omega_1)$,
highlighting the 
dependence on the deformation parameter $b_{\phi_3}$, while the right panel shows the sensitivity
of the function $\bar{\omega}_0\varphi_2(\bar{\omega}_2)$ to $c_{\phi_3}$.
In both cases, variations of the parameters lead to moderate distortions of the 
distribution, affecting both the position and the width of the peak, while preserving 
the overall normalisation and the expected exponential fall-off.

For each DLCDA, we compute the first moments as defined in~\refeq{def_moments}, which read
\begin{align}
    \langle \omega_1\rangle_{[F]} &= 
    \omega_0 \, N_{[F]} \, \frac{(1+k)\big(1+(2+k) \,b_{[f]}\big)}{1+(1+k) \, b_{[f]}} \,, \qquad 
     \langle \bar{\omega}_2\rangle_{[F]} = 
    \bar\omega_0 \, N_{[F]} \, \frac{(1+\bar k)\big(1+(2+\bar k) \,c_{[f]}\big)}{1+(1+\bar k) \, c_{[f]}} \,.
\end{align}
The moments are proportional to the characteristic scales $\omega_0$ ($\bar\omega_0$)  and to the normalisation $N_{[F]}$, 
with additional dependence on the shape parameters $b_{[f]}$ (for $\omega_1$) and $c_{[f]}$ (for $\bar{\omega}_2$). 
Inverting these relations, one can express the model parameters in terms of the first moments. 
These, in turn, are not independent but are constrained by the relations derived in 
\refeqs{w1_moment_relations-Psi4}{w2_moment_relations-Phi6}, and can be written entirely in terms of 
$\lambda_E^2$, $\lambda_H^2$, $\bar{\Lambda}$, and the four independent moments 
$\langle \omega_1 \rangle_{\Phi_3}$, $\langle \omega_1 \rangle_{\Phi_4}$, 
$\langle \omega_1 \rangle_{\Psi_4}$, and $\langle \omega_1 \rangle_{\Phi_5}$.

\section{Implementing the radiative tail of \texorpdfstring{$\boldsymbol{\phi_3}$}{}}
\label{sec:radiative_tail}

The relations derived in the previous section between the matrix elements of \emph{local}
operators and the \emph{moments} of the DLCDAs no longer hold 
if radiative corrections to the non-local HQET operators
are taken into account~\cite{Lange:2003ff,Lee:2005gza}. This can be traced back to the 
non-trivial contributions coming from the Wilson lines
connecting the light fields with the heavy-quark field.
As a consequence, on general grounds, the short-distance expansion of the (D)LCDAs (see, e.g., Ref.~\cite{Kawamura:2008vq}) is given in terms of the same kind of matrix elements 
of local operators as considered above, but multiplied with Wilson coefficients that --- at one-loop order in the $\overline{\textrm{MS}}$ scheme --- depend 
logarithmically on the light-cone coordinates via 
\begin{align}
L_1 = \ln(i z_1 \hat\mu) \,, 
\qquad 
L_2 = \ln(i z_2 \hat\mu )\,, \qquad 
\mbox{with $\hat\mu= \mu \, e^{\gamma_E}$}
\,. 
\end{align}
Translated to momentum space, this generates the so-called ``radiative tails'' with only a power-like fall-off at large light-cone momenta.
The presence of these radiative tails makes the normalisation and the positive moments of (D)LCDAs ill-defined in HQET.
On the other hand, for short distances $|z_i| \sim {\cal O}(1/\mu) \ll 1/\Lambda_{\rm QCD}$, the radiative tail can be calculated explicitly
in fixed-order perturbation theory and can still be used as a
constraint on the (D)LCDA models.
In a recent paper~\cite{Feldmann:2025dcs}, two of us have 
proposed a convenient method to combine the perturbative 
result for the radiative tail with a generic parametrisation 
of the LCDA at large distances. While in that paper we considered 
the (conventional) three-particle LCDAs of the $\Lambda_b$ baryon,
we now apply the same method to the leading-twist three-particle DLCDA of the $B$-meson.

The starting point is the one-loop RGE
for $\Phi_3(z_1,z_2;\mu)$ in position space~\cite{BenekeBoeer:2026},
which can be derived from the corresponding momentum-space
expressions in Refs.~\cite{Huang:2023jdu,Bartocci:2024bbf},
\begin{align}
\begin{aligned}
    \frac{d}{d \ln \mu} \, \Phi_3(z_{1},z_{2};\mu)&= 
    \int_0^{1^+} dt \, G_1(t,z_1;\mu) \, \Phi_3(tz_1,z_2;\mu) 
    \\ & \qquad {} +
    \int_0^{1^+} dt \, G_2(t,z_2;\mu) \, \Phi_3(z_1,tz_2;\mu) 
    \\[0.25em] & \qquad \quad  {} + 
    g_3(z_1,z_2;\mu) \, \Phi_3(z_1,z_2;\mu)\,,
\end{aligned}
\end{align}
with
\begin{align} 
 G_1(t,z_1;\mu) &=    
  -\frac{\alpha_{s} C_{F}}{\pi} 
  \left\{    \left(
     \ln\left(i(z_{1}-i0^{+})\hat\mu\right)+\frac{1}{2}\right)
     \delta(1-t) 
  -  
    \left[\frac{t}{1-t}\right]_{+} \right\} \,,
\\
G_2(t,z_2;\mu) &=     - t \, \frac{\alpha_{s} C_{A}}{\pi} \left\{ 
    \left(\ln\left(i(z_{2}-i0^{+})\hat\mu\right) 
    +\frac{1}{2}\right) \delta(1-t) 
    -  \left[\frac{t}{1-t}\right]_{+} 
     \right\} \,.
\end{align}
Here the term in curly brackets is the position-space version of the Lange-Neubert kernel~\cite{Braun:2003wx},
and 
\begin{align}
    g_3(z_1,z_2;\mu) &=  -\frac{\alpha_{s} C_{A}}{\pi} \bigg(\frac{1+i\pi}{2}-i\pi \,\theta(z_{1}) \, \theta(z_{2})\bigg) \,.
\end{align}
Notice that the latter 
contribution reflects the (partonic) final-state interactions, which lead to  a (perturbatively generated) imaginary part.
As explained in Ref.~\cite{BenekeBoeer:2026}, in the context of factorisation theorems, the
hard-scattering kernels do not support the region where $z_{1,2}<0$, and therefore
we set $\theta(z_1) \theta(z_2) = 1$ in the following. 
In momentum space this is reflected by restricting ourselves to positive light-cone momenta $\omega_1,\bar{\omega}_2\geq 0$. 
This ``reduced RGE''~\cite{Beneke:2022msp} therefore also preserves the factorised form of the two light-cone directions (see \refeq{fmodel})~\cite{Bartocci:2024bbf}.

The RGE can be solved iteratively, writing
\begin{align}
    \Phi_3(z_1,z_2;\mu) & = \Phi_3^{(0)}(z_1,z_2) 
    + \frac{\alpha_s}{4\pi} \, \Phi_3^{(1)}(z_1,z_2;\mu) + {\cal O}(\alpha_s^2)\,,
\end{align}
such that the function $\Phi_3^{(0)}(z_1,z_2)$ can be interpreted
as a tree-level model/parametrisation for the DLCDA. As such,
the function $\Phi_3^{(0)}(0,0)$ and all its derivatives are finite,
and thus the normalisation and all the positive moments in momentum space do exist.
On the other hand, plugging the above ansatz into the RGE yields
a solution for the one-loop correction $\Phi_3^{(1)}(z_1,z_2;\mu)$.
Decomposing the solution according to the kernels $G_1$, $G_2$ and
$g_3$, one obtains
\begin{align}
    \Phi_3^{(1)}(z_1,z_2;\mu) \Big|_{G_1}
    &= -C_F \left( \left( 4 L_1 -2 \ln \frac{\mu}{\mu_1(z_1)} + 2
    \right)  \ln \frac{\mu}{\mu_1(z_1)}
    + c_1(z_1) \right) \Phi_3^{(0)}(z_1,z_2)
    \nonumber\\*
    & \quad {} + C_F 
    \left( 4 \ln \frac{\mu}{\mu_1(z_1)} + c_2(z_1) \right) \int_0^1 dt \left[ \frac{t}{1-t}\right]_+ \Phi_3^{(0)}(t z_1,z_2) \,,
    \label{eq:G1contr}
    \\ 
    \Phi_3^{(1)}(z_1,z_2;\mu) \Big|_{G_2} 
    & = - C_A \,  \left(
    \left( 4 L_2 -2 \ln \frac{\mu}{\mu_2(z_2)} + 2
    \right)  \ln \frac{\mu}{\mu_2(z_2)}
    + c_3(z_2) 
    \right) \Phi_3^{(0)}(z_1,z_2)
    \nonumber\\*
    & \quad {} + C_A
    \left( 4 \ln \frac{\mu}{\mu_2(z_2)} + c_4(z_2) \right) \int_0^1 dt \left[ \frac{t}{1-t}\right]_+ t \, \Phi_3^{(0)}(z_1,t z_2)\,,
    \label{eq:G2contr}
\end{align}
and 
\begin{align} 
    \Phi_3^{(1)}(z_1,z_2;\mu) \Big|_{g_3} 
    & =  - C_A \, \frac{1-i\pi}{2} \left( 4 \ln \frac{\mu}{\mu_3(z_1,z_2)} + c_5(z_1,z_2) \right) \, \Phi_3^{(0)}(z_1,z_2) \,.
\end{align}  
Here we have introduced a number of $z_i$-dependent integration constants, which 
appear as reference scales $\mu_i$ in logarithms or constants $c_i$ multiplying the 
tree-level input function. Notice that the choice of the integration constants refers to a definition of a renormalisation scheme for the input function $\Phi_3^{(0)}$ which thus implicitly depends on the integration constants. The meaning of this shall become clearer from below.

\subsection{Short-distance limit}

Expanding around the limit $z_i \to 0$, the above result corresponds 
to the so-called ``radiative tail'' of the DLCDA, which could be obtained
from a partonic one-loop matching calculation for the short-distance expansion of
the respective di-light-cone operator. Here, the contribution from the kernel $G_1$
is the same as for the standard two-particle LCDA, leading to~\cite{Feldmann:2025dcs}
\begin{align}
    \Phi_3^{(1)}(z_1,z_2;\mu) \Big|_{G_1}
    &= -C_F \left( \left( 2 L_1 + 2
    \right)  L_1 
    + \frac{5\pi^2}{12} \right) \left( N_{\Phi_3} -i z_1 \,  \langle \omega_1 \rangle_{\Phi_3} -i z_2 \, \langle \bar{\omega}_2 \rangle_{\Phi_3} + \ldots \right) 
    \nonumber\\*
    & \quad {} + C_F 
    \left( 4 L_1 + c_2 \right)
    \left(\frac{i}{2} \,  z_1 \, \langle \omega_1 \rangle_{\Phi_3} + \ldots \right)\,,
\end{align}
where we have used that the reference scale 
can only depend on $z_1$ in the form $\mu_1(z_1):= e^{-\gamma_E}/iz_1$, and $c_1(z_1) = \frac{5\pi^2}{12}$ and $c_2=\text{const.}$ on general grounds. The expansion of the input function on the right-hand side in this case is determined by the local operators which define the (tree-level) moments discussed above,
where the derivatives from the Taylor expansion in position space translate to the corresponding moments in momentum space.
The contribution of the kernel $G_2$ can be treated in a similar way,
\begin{align}
    \Phi_3^{(1)}(z_1,z_2;\mu) \Big|_{G_2}
    &= -C_A \left( \left( 2 L_2 + 2
    \right)  L_2 
    + \frac{5\pi^2}{12} \right) \left( N_{\Phi_3} -i z_1 \,  \langle \omega_1 \rangle_{\Phi_3} -i z_2 \, \langle \bar{\omega}_2 \rangle_{\Phi_3} + \ldots \right) 
    \cr 
    & \quad {} + C_A 
    \left( 4 L_2 + c_4 \right)  \left( -\frac12 \, N_{\Phi_3} + \frac{i}{2}\, z_1 \,  \langle \omega_1 \rangle_{\Phi_3} + \frac{5i}{6} \, z_2 \, \langle \bar{\omega}_2 \rangle_{\Phi_3} + \ldots \right) \,.
\end{align}
The contribution from the kernel $g_3$ requires closer inspection of the one-loop diagrams contributing to the anomalous-dimension kernels.
From this we infer that the non-trivial analytic structure only involves the gluon separation $z_2$,
such that 
\begin{align} 
    \Phi_3^{(1)}(z_1,z_2;\mu) \Big|_{g_3} 
    & = - C_A \, \frac{1-i\pi}{2} \left( 
    4 L_2 + c_5\right) \left( N_{\Phi_3} -i z_1 \,  \langle \omega_1 \rangle_{\Phi_3} -i z_2 \, \langle \bar{\omega}_2 \rangle_{\Phi_3} + \ldots \right) \,,
\end{align} 
which corresponds to setting $\mu_3(z_1,z_2)=\mu_2(z_2) \to e^{-\gamma_E}/iz_2$.

\subsection{Interpolating with hadronic input function}

The interpolation of the short-distance limit with a generic hadronic model
for finite $z_{1,2}$ can be achieved by modifying the integration constants,
as explained in Ref.~\cite{Feldmann:2025dcs}. Our default choice amounts to setting
\begin{align}
    \hat\mu_1(z_1) & := \frac{1 +i \hat\mu_F  z_1}{i z_1} \,, 
    \qquad 
    \hat\mu_2(z_2) := \frac{1 +i \hat\mu_F  z_2}{iz_2} \,, 
\end{align}
together with 
\begin{align}
\begin{aligned}
    c_1(z_1) &= c_3(z_2)  = \frac{5\pi^2}{12} \,, 
    & \qquad\qquad
    c_2(z_1) &= \frac{c_2}{1+i \hat\mu_F z_1} \,, 
    \\
    c_4(z_2) &= c_4 \, \frac{1+ 2 i \hat\mu_F z_2 }{(1+i \hat\mu_F z_2)^2} \,, 
    &
    c_5(z_1,z_2) &= c_5 \,.
\end{aligned}
\end{align}
Here, the reference scale $\mu_F$ is introduced in such a way that the \emph{shape} of the DLCDA at \emph{large} values of $|z_i|$ approaches that of the input function $\Phi_3^{(0)}$.
It should be emphasised that the particular choice for the integration constants preserves the analytic properties of the DLCDAs in position space.\footnote{Here we restrict ourselves again to the situation where $\omega_1$ and $\bar{\omega}_2$ have only positive support. For a more general discussion of the analytic properties of the DLCDAs after including renormalisation-group effects, see \cite{Huang:2023jdu} and the supplementary material therein.}
Apart from this, our choice remains to some extent arbitrary and should be considered as part of the modelling,
which is reflected by the numerical dependence on the auxiliary scale $\mu_F$.

The previous result, associated with the perturbative matching calculation for the radiative tail, on the other hand, would 
be recovered by setting $\mu_F \to 0$. 
For finite $\mu_F$, we also have to consider a modified input function,
where the short-distance expansion now reads
\begin{align}
    \Phi_3^{(0)}(z_1,z_2; \mu_F) &= 
    N_{\Phi_3}(\mu_F) - i z_1 \, \langle \omega_1 \rangle_{\Phi_3}(\mu_F) - i z_2 \, \langle \bar{\omega}_2 \rangle_{\Phi_3}(\mu_F) + \ldots \,,
\end{align}
i.e.\ all moments now implicitly depend on $\mu_F$ as well.
Comparing the short-distance expansion in one or the other case, we obtain a relation between the corresponding moments, which for our parametrisation reads:
\begin{align}
\begin{aligned}
    N_{\Phi_3}(\mu_F) &= N_{\Phi_3}(0) \,, 
    \\ 
    \langle \omega_1 \rangle_{\Phi_3}(\mu_F) &= 
    \langle \omega_1 \rangle_{\Phi_3}(0) 
    + \frac{\alpha_s C_F}{2\pi} 
    \, \hat\mu_F  \, N_{\Phi_3}(0) + \ldots \,,
    \\ 
     \langle \bar{\omega}_2 \rangle_{\Phi_3}(\mu_F) &= 
    \langle \bar{\omega}_2 \rangle_{\Phi_3}(0) 
    + 
  \frac{\alpha_s C_A}{2\pi} \, 
  (3-i\pi)
  \,\hat\mu_F  \, N_{\Phi_3}(0) + \ldots \,.
 \label{eq:momrel}
\end{aligned}
\end{align}
Within our parametrisation this can be realised
by adjusting the shape parameters as follows
\begin{align}
    b_{\phi_3}(\mu_F) &= b_{\phi_3}(0)+(1+2b_{\phi_3}(0))^2  \,\frac{\hat\mu_F}{2\omega_0} \, \frac{\alpha_s C_F}{2\pi} + {\cal O}(\alpha_s^2) \,, 
    \\
    c_{\phi_3}(\mu_F) &= c_{\phi_3}(0)+(1+3c_{\phi_3}(0))^2  \,\frac{\hat\mu_F}{3\bar\omega_0} \, \frac{\alpha_s C_A}{2\pi} \, (3-i\pi)+ {\cal O}(\alpha_s^2) \,. 
\end{align}
Notice that the perturbative shift in $\langle\bar{\omega}_2\rangle$ is enhanced by a factor $C_A/C_F$ compared to $\langle \omega_1\rangle$, which reflects the different renormalisation-group effects on the shape of the DLCDAs in one or the other direction. As already noticed in the previous section, this also suggests to consider $\langle \omega_1\rangle$ and $\langle\bar{\omega}_2\rangle$ to be controlled by different reference scales $\bar \omega_0 > \omega_0$ as well.

The moments on the left-hand side of \refeq{momrel} then define
the parametrisation of the input function.
The position-space parametrisation 
for our choice in \refeq{fmodel} can be obtained
from a Laplace transform 
from $(\omega_1,\bar{\omega}_2) \to (i z_1,i z_2)$, which leads to a rational function,
\begin{align}
    \Phi_3^{(0)}(z_1,z_2;\mu_F) &= N_{\Phi_3}(\mu_F) \,
    \frac{(1 + 2 b_{\phi_3}(\mu_F) + i\omega_0 z_1) \, (1 + 3 c_{\phi_3}(\mu_F) + i\bar\omega_0 z_2)}{(1 + 2 b_{\phi_3}(\mu_F)) \, (1 + 3 c_{\phi_3}(\mu_F)) \, (1 + 
    i \omega_0 z_1)^3 (1 + i\bar \omega_0 z_2)^4}
    \,.
\end{align} 
With this, we can easily obtain the one-loop correction. 
For this we need the integrals 
\begin{align}
  \Phi_3^{\rm aux,1}(z_1,z_2;\mu_F) & :=  \int_0^1 dt \left[\frac{t}{1-t}\right]_+ \Phi_3^{(0)}(tz_1,z_2;\mu_F) \nonumber
    \\    & = \left( \ln(1+i\omega_0 z_1) + \frac{i \omega_0 z_1 \, b_{\phi_3}(\mu_F)}{1 + 2 b_{\phi_3}(\mu_F) + i\omega_0 z_1} \right) \Phi_3^{(0)}(z_1,z_2;\mu_F)\,,
    \label{eq:aux_pos1}
    \\
    \Phi_3^{\rm aux,2}(z_1,z_2;\mu_F) & :=  \int_0^1 dt \left[\frac{t}{1-t}\right]_+ 
    t \, \Phi_3^{(0)}(z_1,t z_2;\mu_F) \nonumber
    \\
    & = \left( \ln(1+i\bar\omega_0 z_2) 
    -\frac{1 + i \bar\omega_0 z_2 + (3  - 2i \bar\omega_0 z_2)  \, c_{\phi_3}(\mu_F)}{2 + 6 c_{\phi_3}(\mu_F) + 2 i\bar\omega_0 z_2}  \right) \Phi_3^{(0)}(z_1,z_2;\mu_F)
    \,,
    \label{eq:aux_pos2}
\end{align}
appearing in \refeqa{G1contr}{G2contr}.

\subsubsection{Translation to momentum space}

The result for $\Phi_3^{(1)}(z_1,z_2)$ in position space can easily be translated to momentum space. Here and in the following, all functions and input parameters are understood to depend on the reference scale $\mu_F$. For our form of the input function it is convenient to exploit the fact that the result takes a factorised form (up to corrections of order $\alpha_s^2$), i.e.
\begin{align}
\begin{aligned}
    \phi_3(\omega_1,\bar{\omega}_2) & \simeq N_{\Phi_3} \, \varphi_1(\omega_1) \, \varphi_2(\bar{\omega}_2) 
    \\
    &= N_{\Phi_3} \left( \varphi_1^{(0)}(\omega_1) + \frac{\alpha_s}{4\pi} \, \varphi_1^{(1)}(\omega_1) +\ldots \right) 
    \left( \varphi_2^{(0)}(\bar{\omega}_2) + \frac{\alpha_s}{4\pi} \, \varphi_2^{(1)}(\bar{\omega}_2) +\ldots \right)
    \,.
    \label{eq:Phi3pert1}
\end{aligned}
\end{align}
Notice that the analytic structure of our setup restricts the support region in momentum space to $\omega_1\geq 0$ and $\bar{\omega}_2\geq 0$. 
Furthermore, all terms in the above expression implicitly or explicitly depend on the auxiliary scale $\mu_F$.
Using the same kind of relations as in Ref.~\cite{Feldmann:2025dcs},
the result for the individual one-loop terms can be written in terms of convolution integrals,
\begin{align}
    \varphi_1^{(1)}(\omega_1) &=
    \int\limits_0^{\omega_1} d\omega_1' 
    \left( K_{1a}(\omega_1-\omega_1') \,  \varphi_1^{(0)}(\omega_1') 
    + K_{1b}(\omega_1-\omega_1') \,  \frac{d\varphi_1^{(0)}(\omega_1')}{d\omega_1'}
    + K_{1c}(\omega_1-\omega_1') \, \varphi_1^{\rm aux}(\omega_1') \right) \,,
    \label{eq:Phi3pert2}
\end{align}
and analogously for $\varphi_2^{(1)}(\bar{\omega}_2)$.
Here we find it convenient
to decompose the individual contributions
in terms of auxiliary expressions that can
be easily derived for a given input function.
In our case this reads
\begin{align}
\begin{aligned}
    \frac{d\varphi_1^{(0)}(\omega_1)}{d\omega_1} 
    &= \frac{1}{\omega_0}
    \left(-1 + \frac{\omega_0}{\omega_1} + \frac{b_{\phi_3}}{1+b_{\phi_3}\, \omega_1/\omega_0} \right)\varphi_1^{(0)}(\omega_1) \,,
    \\
    \frac{d\varphi_2^{(0)}(\bar{\omega}_2)}{d\bar{\omega}_2} &= 
    \frac{1}{\bar\omega_0} \left(-1 + \frac{2\bar\omega_0}{\bar{\omega}_2} + \frac{c_{\phi_3}}{1+c_{\phi_3} \, \bar{\omega}_2/\bar\omega_0} \right)\varphi_2^{(0)}(\bar{\omega}_2) \,, 
\end{aligned}
\end{align}
as well as
\begin{align}
\begin{aligned}
    \varphi_1^{\rm aux}(\omega_1) & := \varphi_1^{(0)}(\omega_1) + \int_{\omega_1}^\infty d\omega_1' \left[ \frac{\omega_1}{\omega_1'(\omega_1'-\omega_1)}\right]_+ \varphi_1^{(0)}(\omega_1') 
    \\
    &=  \left( 1+ \ln \frac{\omega_0}{\omega_1} -\gamma_E + \frac{b_{\phi_3}}{1+b_{\phi_3} \, \omega_1/\omega_0} \right)\varphi_1^{(0)}(\omega_1) \,,
\end{aligned}
\end{align}
and 
\begin{align}
\begin{aligned}
    \varphi_2^{\rm aux}(\bar{\omega}_2) & :=  \int_{\bar{\omega}_2}^\infty d\bar{\omega}_2' \left[ \frac{\bar{\omega}_2^2}{(\bar{\omega}_2')^2 (\bar{\omega}_2'-\bar{\omega}_2)}\right]_+ \varphi_2^{(0)}(\bar{\omega}_2') 
    \\    
    &=  \left(1 + \ln \frac{\bar\omega_0}{\bar{\omega}_2} -\gamma_E + \frac{c_{\phi_3}}{1+c_{\phi_3}\, \bar{\omega}_2/\bar\omega_0} \right)\varphi_2^{(0)}(\bar{\omega}_2) \,,
\end{aligned}
\end{align}
where the Laplace transform of the functions $\varphi_{1,2}^{\rm aux}$ reproduces the expressions in \refeqa{aux_pos1}{aux_pos2}.
Using the formulas for the inverse Laplace transform as given in \refapp{laplace}, the convolution kernels can be derived as\footnote{Notice that the representation of the convolution kernels is not unique, as terms can be moved from one kernel to the other by partial-integration.}
\begin{align}
    K_{1a}(\omega) &= 
    C_F \left( 2 \, \ln^2\frac{\mu_F}{\omega} - \frac{\pi^2}{3} 
    \right) \frac{e^{-\omega/\hat\mu_F}}{\hat\mu_F}  
  + C_F 
   \left(  -2 \, \frac{e^{-\omega/\hat\mu_F}-1}{\omega} \right) \,,
\end{align}
and 
\begin{align}
    K_{1b}(\omega) & = C_F \left( - 2 \ln^2 \frac{\mu}{\omega}  -2 \, \ln \frac{\mu}{\mu_F} - \frac{\pi^2}{12} 
    \right)
    + C_F \left( 2 \ln^2 \frac{\mu_F}{\omega} - \frac{\pi^2}{3} \right) e^{-\omega/\hat\mu_F}\,,
\end{align}
and 
\begin{align}
    K_{1c}(\omega) &= C_F \left( 
      4 \, \ln \frac{\mu}{\mu_F} \, \delta(\omega) 
      + 4 \, \frac{e^{-\omega/\hat\mu_F}-1}{\omega} 
      + c_2 \, \frac{e^{-\omega/\hat\mu_F}}{\hat \mu_F}\right)\,. 
\end{align}
The analogous expressions for the kernels contributing to $\varphi_2^{(1)}(\bar{\omega}_2)$ read
\begin{align}
&K_{2a}(\omega)=\frac{C_A}{C_F} \, K_{1a}(\omega) 
  - C_{A} \, \frac{1-i\pi}{2} \left( 4 \, \frac{e^{-\omega/\hat\mu_F}-1}{\omega} \right) \,, 
\\
&K_{2b}(\omega)=\frac{C_A}{C_F} \, K_{1b}(\omega)
  - C_{A} \, \frac{1-i\pi}{2} \left( 4 \, \ln \frac{\mu}{\mu_F}+ c_5 \right) \,, 
\\ 
& K_{2c}(\omega)= C_A \left( 
      4 \, \ln \frac{\mu}{\mu_F} \, \delta(\omega) 
      + 4 \, \frac{e^{-\omega/\hat\mu_F}-1}{\omega} 
      + c_4 \left(2 -\frac{\omega}{\hat\mu_F}\right)
      \frac{e^{-\omega/\hat\mu_F}}{\hat \mu_F}\right) \,.
\end{align}

It is instructive to consider the region of large light-cone momenta $\omega_1,\bar{\omega}_2 \gg \hat \mu_F$ which is usually referred to as the radiative tail.
Exploiting the fact that the input functions only have significant support for small values of $\omega'$, the above-defined kernels in that limit then reduce to 
\begin{align}
    K_{1a}(\omega-\omega') & \to 
C_F 
   \left(\frac{2}{\omega} + \frac{2\omega'}{\omega^2} + \ldots \right) 
\,,
\\
    K_{1b}(\omega-\omega') & \to C_F \left( - 2 \ln^2 \frac{\mu}{\omega}  
    -4 \, \frac{\omega'}{\omega} \, \ln \frac{\mu}{\omega}
    -2 \, \frac{\omega'{}^2}{\omega^2}\left( 1 + \ln \frac{\mu}{\omega}\right) +\ldots 
    \right) 
    \,, 
    \\
    K_{1c}(\omega- \omega') &\to C_F \left( 
  - \frac{4}{\omega} - \frac{4\omega'}{\omega^2} + \ldots
   \right) \,,
\end{align}
and similarly,
\begin{align}
    K_{2a}(\omega-\omega') & \to 
    (2-i\pi)
    \,\frac{C_A}{C_F} K_{1a}(\omega-\omega') 
    \,, 
    \\ 
    K_{2b}(\omega-\omega') & \to \frac{C_A}{C_F} \, K_{1b}(\omega-\omega') \,, 
    \\ 
    K_{2c}(\omega-\omega') & \to \frac{C_A}{C_F} \, K_{1c}(\omega-\omega') \,.
\end{align}
where the constant term in $K_{1b,2b}$ can be neglected because it vanishes in the convolution integral.
With this, the $\omega'$-integrals can be expressed in terms of the moments of the input function again.
Using that 
\begin{align}
 \int_0^\infty d\omega_1' \left\{ 1,\omega_1',\omega_1'^2 \right\} \frac{d\varphi_1^{(0)}(\omega_1')}{d\omega_1'} &=
 \left\{ 0 , -1 , - 2 \, \frac{\langle \omega_1\rangle_{\Phi_3}}{N_{\Phi_3}} \right\} \,,
 \cr 
 \int_0^\infty d\omega_1' \left\{ 1,\omega_1' \right\} \varphi_1^{\rm aux}(\omega_1') &= \left\{ 0 , -\frac12 \, \frac{\langle \omega_1\rangle_{\Phi_3}}{N_{\Phi_3}} \right\} \,,
\end{align}
the final result for the radiative tail in the $\omega_1$ direction takes the form
\begin{align}
\begin{aligned}
    \varphi_1(\omega_1) \big|_{\rm tail} &= 
    \frac{\alpha_s C_{F}}{4\pi} 
    \left( 4 \ln \frac{\mu}{\omega_1} + 2 \right) \frac{1}{\omega_1} 
    \\ 
    & \quad {} + \frac{\alpha_s C_{F}}{4\pi} \left( 
    4 \ln \frac{\mu}{\omega_1} + 8\right) \frac{1}{\omega_1^2} \, \frac{\langle \omega_1 \rangle_{\Phi_3}}{N_{\Phi_3}} 
    + \ldots \,.
    \label{eq:Phi3tail}
\end{aligned}
\end{align}
Similarly, with 
\begin{align}
 \int_0^\infty d\bar{\omega}_2' \left\{ 1,\bar{\omega}_2',\bar{\omega}_2'^2 \right\} \frac{d\varphi_2^{(0)}(\bar{\omega}_2')}{d\bar{\omega}_2'} &=
 \left\{ 0 , -1 , - 2 \, \frac{\langle \bar{\omega}_2\rangle_{\Phi_3}}{N_{\Phi_3}} \right\} \,,
 \cr 
 \int_0^\infty d\bar{\omega}_2' \left\{ 1,\bar{\omega}_2' \right\} \varphi_2^{\rm aux}(\bar{\omega}_2') &= \left\{ -\frac12 , -\frac56 \, \frac{\langle \bar{\omega}_2\rangle_{\Phi_3}}{N_{\Phi_3}} \right\} \,,
\end{align}
we find
\begin{align}
\begin{aligned}
    \varphi_2(\bar{\omega}_2) \big|_{\rm tail} &= 
    \frac{\alpha_s C_{A}}{4\pi} 
    \left( 4  \ln \frac{\mu}{\bar{\omega}_2} + { (6-2i\pi)}\right) \frac{1}{\bar{\omega}_2} 
    \\ 
    & \quad {} + \frac{\alpha_s C_{A}}{4\pi} \left( 
    4 \ln \frac{\mu}{\bar{\omega}_2} 
    { + \left( \frac{34}{3} - 2i\pi \right)}
    \right)
    \frac{1}{\bar{\omega}_2^2} \, \frac{\langle \bar{\omega}_2 \rangle_{\Phi_3}}{N_{\Phi_3}} 
    + \ldots \,.
    \label{eq:Phi3tail2}
\end{aligned}
\end{align}

\subsubsection{Numerical illustration}

\begin{figure}[p]
   \begin{center}
     \includegraphics[width=0.47\linewidth]{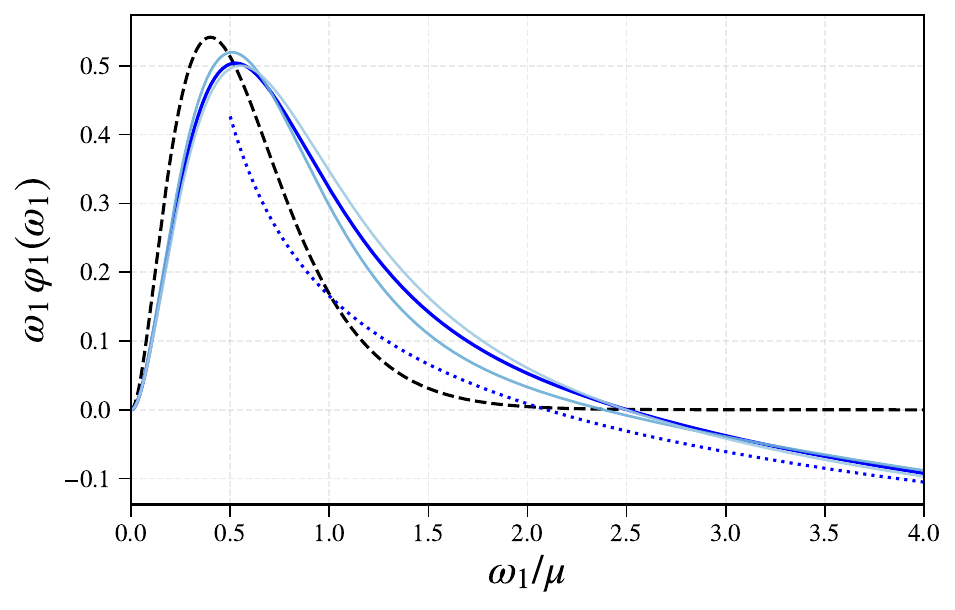}
    \end{center}
    \caption{Effect of the radiative tail in the function $\varphi_1(\omega_1)$: The dashed black line shows the tree-level parametrisation $\varphi_1^{(0)}(\omega_1)$ with the shape parameter $b_{\phi_3}(\mu_F=0)=0$. The blue dotted line shows the radiative tail
    up to terms of order $1/\omega_1^2$, and the solid blue line shows the result of the one-loop improved model. 
    For the numerical illustration we considered the following input parameters: $\omega_0=0.3$~GeV, $\mu=1.5~$GeV, $\alpha_s(\mu)=0.3$, and set the unknown integration constant $c_2=0$ for simplicity. 
    The default value for the auxiliary scale $\mu_F$
    has been set to $1~$GeV, while the light blue lines 
    refer to changing $\mu_F$ to $0.7$~GeV or $1.4$~GeV, respectively. In each case the shape parameter is adjusted accordingly, $b_{\phi_3}(\mu_F) = \frac{\alpha_s C_F}{4\pi} \, \frac{\hat\mu_F}{\omega_0}$.
    Notice that in order to emphasise the radiative tail, we have plotted the product $\omega_1 \, \varphi_1(\omega_1)$ 
    as a function of the dimensionless ratio $\omega_1/\mu$.
  }
    \label{fig:phi1tail}
    \vspace{1.0cm}
   \begin{center}
    \includegraphics[width=0.47\linewidth]{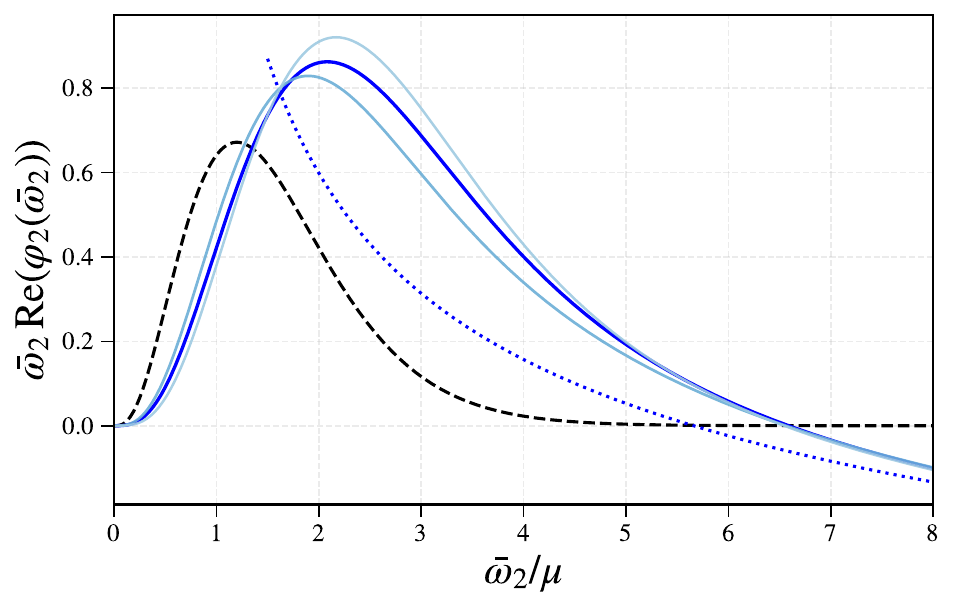}
    \quad 
    \includegraphics[width=0.47\linewidth]{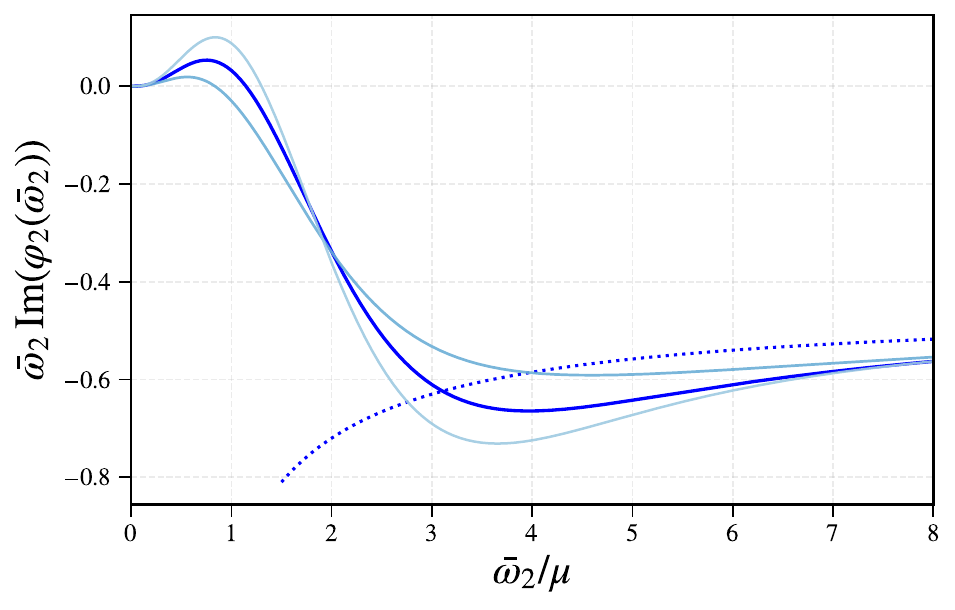}
    \end{center}
   
    \caption{Effect of the radiative tail in the function $\varphi_2(\bar{\omega}_2)$: The dashed black line shows the tree-level parametrisation $\varphi_2^{(0)}(\bar{\omega}_2)$ with the shape parameter $c_{\phi_3}(\mu_F=0)=0$. The blue dotted line shows the radiative tail
    up to terms of order $1/\bar{\omega}_2^2$, and the solid blue line shows the result of the one-loop improved model. 
    For the numerical illustration we considered the following input parameters: $\bar\omega_0=0.6$~GeV, $\mu=1.5~$GeV, $\alpha_s(\mu)=0.3$, and set the unknown integration constants $c_4=c_5=0$ for simplicity. 
    The default value for the auxiliary scale $\mu_F$
    has been set to $1~$GeV, while the light blue lines 
    refer to changing $\mu_F$ to $0.7$~GeV or $1.4$~GeV, respectively. In each case the shape parameter is adjusted accordingly, $c_{\phi_3}(\mu_F) = \frac{\alpha_s C_A}{6\pi} \, (3-i\pi) \,  \frac{\hat\mu_F}{\bar\omega_0}$.
    Notice that in order to emphasise the radiative tail, we have plotted the product $\bar{\omega}_2 \, \varphi_2(\bar{\omega}_2)$ 
    as a function of the dimensionless ratio $\bar{\omega}_2/\mu$. 
    The left panel shows the real part, while the right panel shows the imaginary part (which is absent in $\varphi_2^{(0)}(\bar{\omega}_2)$ by assumption).
    }
    \label{fig:phi2tail}
\end{figure}

To illustrate the numerical effect of including the radiative tail
in the one-loop improved modelling of the DLCDAs, we compare in \reffig{phi1tail} our result for the function $\varphi_1(\omega_1)$ defined in \refeqa{Phi3pert1}{Phi3pert2} 
to a given (tree-level) 
input function $\varphi_1^{(0)}(\omega_1)$ and 
to the radiative tail as given in \refeq{Phi3tail}. 
We stress that the numerical values for the input parameters
used in that figure are chosen for illustrative purposes only. A full-fledged quantitative analysis would require additional input, in particular on the first moments $\langle \omega_1\rangle_{\Phi_3}$ entering the short-distance expansion. Nevertheless, on a qualitative level,
we observe that the implementation of the radiative tail leads to 
a significant shift of the distribution towards larger values of $\omega_1$ and to a zero crossing around $\omega_1 \sim 2.5\mu$. The precise numerical effect at intermediate values of $\omega_1$ depends on the chosen value of the auxiliary scale $\mu_F$, which is also illustrated in that figure.

An analogous plot for the function $\varphi_2(\bar{\omega}_2)$ is presented in \reffig{phi2tail}.
As already pointed out above, the different colour factor in the radiative corrections for the function $\varphi_2^{(1)}(\bar{\omega}_2)$ enhances the effect of the radiative tail, which leads to a broader distribution and a zero crossing (in the real part) at larger values of $\bar{\omega}_2 \sim 6\mu$.
Notice that in that case, the one-loop improvement also leads to a sizeable $\bar{\omega}_2$-dependent imaginary part.

\section{Summary and conclusion}
\label{sec:concl}

In this work we have initiated a systematic study of three-particle di-light-cone distribution amplitudes (DLCDAs) of the $B$-meson in HQET. These can 
be viewed as a generalisation of the conventional $B$-meson light-cone distribution amplitudes, by defining $B$-meson--to--vacuum matrix elements of trilocal HQET operators with the light-antiquark and the gluon field-strength tensor separated along two back-to-back light-like directions.
The DLCDAs are needed to parametrise non-perturbative hadronic effects as they enter, for instance, non-factorisable soft-gluon contributions to rare and non-leptonic exclusive $B$-meson decays.

Starting from the most general Lorentz decomposition, we have shown that the three-particle $B$-meson DLCDAs can be expressed in terms of eight independent distributions, which we have also organised into a basis of definite twist. Furthermore, we derived constraints on the zeroth and first moments of the DLCDAs from the respective local expansion of the operators at tree level, together with the quark and gluon equations of motion. This allowed us to construct phenomenological momentum-space models that are consistent with the expected low-momentum behaviour from the conformal spin of the respective fields, and that are parametrised in terms of a manageable number of well-defined hadronic parameters.

In the context of factorisation approaches, the most relevant DLCDA is the leading-twist distribution $\Phi_3$ that already appeared in the literature in several different contexts. For this function, we discussed in some detail how a tree-level parametrisation is modified by perturbative short-distance corrections inducing so-called radiative tails at large momenta. 
In contrast to the conventional $B$-meson LCDAs, these effects also lead to complex rescattering phases that, in the perturbative calculation, arise from diagrams that connect fields associated with the two different outgoing light-like directions. 
This new source of strong phases in exclusive $B$-decay observables might be of particular relevance in future phenomenological applications. 

Our work thus provides a first step towards implementing DLCDAs in concrete factorisation formulas for various exclusive $B$-meson decays with large hadronic recoil in two opposite light-cone directions.  
This will allow us to calculate contributions from soft gluons attached, for example, to charm loops in radiative $B$ decays or to energetic quarks in non-leptonic two-body decays, and to assess their impact on some of the observables associated with the so-called ``flavour anomalies''.

\Needspace{12\baselineskip}
\section*{Acknowledgements}

The research of PB was funded by the European Union's Horizon 2020 Research and Innovation Programme under the Marie Sk\l{}odowska-Curie grant agreement No. 101146976.
NG has been funded by the Deutsche Forschungsgemeinschaft (DFG, German Research Foundation) -- Emmy-Noether Grant No. 558599025.
The research of TF and DV was supported by the Deutsche Forschungsgemeinschaft (DFG, German Research Foundation) under grant 396021762 - TRR 257.
The research of TF, DV, and NG was supported by Germany’s Excellence Strategy – Cluster of Excellence ``Color meets Flavor'', EXC 3107
– Project-ID 533766364.
The research of MF has received funding from the Cluster of Excellence PRISMA${}^{++}$ (EXC 2118/2, Project ID 390831469) funded by the German Research Foundation (DFG) within the Germany Excellence Strategy.

\appendix
\addcontentsline{toc}{section}{Appendices}

\newcounter{APP}
\renewcommand{\theAPP}{\Alph{APP}}
\setcounter{APP}{0}
\renewcommand{\theequation}{\theAPP.\arabic{equation}}
\renewcommand{\theHequation}{\theAPP.\arabic{equation}}

\appsection{Lorentz decomposition for the three-particle DLCDAs}
\label{app:genDAs}

In this appendix, we parametrise the matrix element
\begin{align}
    \label{eq:genTME}
    \langle 0| \bar q(x)\, gG_{\mu\nu}(y)\Gamma h_v(0) |\bar B(v)\rangle 
\end{align}
in terms of \emph{generalised} DLCDAs.
Unlike in the rest of the paper, the four-vectors $x$ and $y$ are here taken to be light-like, satisfying $x^2=y^2=0$, but they are not restricted to lie on opposite light-like rays. 
Moreover, as we discuss below, most of the relations derived in this appendix continue to hold for arbitrary four-vectors $x$ and $y$.
As in \refsec{defDDAs}, we follow Ref.~\cite{Geyer:2005fb} and begin by constructing the most general decomposition of
$\langle 0| \bar q(x)\, gG_{\mu\nu}(y)\gamma_5\gamma_\rho\sigma_{\alpha\beta} h_v(0) |\bar B(v)\rangle$.
We find
\begin{align}
    \langle G_{\mu\nu} & \gamma_5 \gamma_\rho \sigma_{\alpha\beta} \rangle
    =
    \nonumber\\*
    &
        g_{\alpha[\mu}\, g_{\nu]\beta}\, g_{\rho\sigma}\, \C^\sigma_1
        +
        g_{\sigma[\mu}\, g_{\nu][\alpha}\, g_{\beta]\rho}\, \C^\sigma_2
        +
        g_{\sigma[\alpha}\, g_{\beta][\mu}\, g_{\nu]\rho}\,\C^\sigma_3
        + g_{\kappa[\mu}\,g_{\nu]\tau}\,
        g_{\lambda[\alpha}\,{g_{\beta]}}^\tau\,g_{\rho\sigma}
        \frac{v^\kappa v^\lambda}{v^2}\,\C^\sigma_6
    \nonumber\\
    &
        +
        \frac{v_{[\mu } x_{\nu]} }{v\sdot x }\,
        g_{\rho[\alpha}\,g_{\beta]\sigma}\,\X^\sigma_4
        + 
        \frac{v_{[\alpha}  x_{\beta]} }{v\sdot x }\,
        g_{\rho[\mu}\, g_{\nu]\sigma}\,\X^\sigma_5
        + g_{\kappa[\mu}\,g_{\nu]\tau}\,
        g_{\lambda[\alpha}\,{g_{\beta]}}^\tau\,g_{\rho\sigma}
        \bigg(
            \frac{v^\kappa x^\lambda}{v\sdot x }\,\X^\sigma_7
            +\frac{x^\kappa v^\lambda}{v\sdot x }\,\X^\sigma_8
        \bigg)
    \nonumber\\
    &
        + g_{\kappa[\mu}\,g_{\nu]\tau}\,
        g_{\lambda[\alpha}\,{g_{\beta]}}^\tau\,g_{\rho\sigma}
        \frac{v^2 x^\kappa x^\lambda}{(v\sdot x )^2}\,\X^\sigma_9
        + 
        \frac{v_{[\mu } x_{\nu]}\,v_{[\alpha} x_{\beta]} }{(v\sdot x )^2}\,
        g_{\rho\sigma}\, \X^\sigma_{10}
    \nonumber\\
    &
        +
        \frac{v_{[\mu } y_{\nu]} }{v\sdot y}\,
        g_{\rho[\alpha}\,g_{\beta]\sigma}\,\Y^\sigma_{11}
        + 
        \frac{v_{[\alpha}  y_{\beta]} }{v\sdot y}\,
        g_{\rho[\mu}\, g_{\nu]\sigma}\,\Y^\sigma_{12}
        + g_{\kappa[\mu}\,g_{\nu]\tau}\,
        g_{\lambda[\alpha}\,{g_{\beta]}}^\tau\,g_{\rho\sigma}
        \bigg(
         \frac{v^\kappa y^\lambda}{v\sdot y}\,\Y^\sigma_{13}
        +\frac{y^\kappa v^\lambda}{v\sdot y}\,\Y^\sigma_{14}
        \bigg)
    \nonumber\\
    &
        + g_{\kappa[\mu}\,g_{\nu]\tau}\,
        g_{\lambda[\alpha}\,{g_{\beta]}}^\tau\,g_{\rho\sigma}
        \frac{v^2 y^\kappa y^\lambda}{(v\sdot y)^2}\,\Y^\sigma_{15}
        + 
        \frac{v_{[\mu } y_{\nu]}\,v_{[\alpha} y_{\beta]} }{(v\sdot y)^2}\,
        g_{\rho\sigma}\, \Y^\sigma_{16}
    \nonumber\\
    &
        +
        \frac{1}{(v\sdot x )(v\sdot y)} \bigg[
        x_{[\mu } y_{\nu]} \,
        g_{\rho[\alpha}\,g_{\beta]\sigma}\,\W^\sigma_{17}
        + 
        x_{[\alpha}  y_{\beta]} \,
        g_{\rho[\mu}\, g_{\nu]\sigma}\,\W^\sigma_{18}
    \nonumber\\
    &
        + g_{\kappa[\mu}\,g_{\nu]\tau}\,
        g_{\lambda[\alpha}\,{g_{\beta]}}^\tau\,g_{\rho\sigma}
        \bigg(
             x^\kappa y^\lambda\,\W^\sigma_{19}
            +y^\kappa x^\lambda\,\W^\sigma_{20}
        \bigg)
        + 
        v_{[\mu } x_{\nu]}\,v_{[\alpha} y_{\beta]} \,
        g_{\rho\sigma}\, \W^\sigma_{21}
        + 
        v_{[\mu } y_{\nu]}\,v_{[\alpha} x_{\beta]} \,
        g_{\rho\sigma}\, \W^\sigma_{22}
        \bigg]
    \nonumber\\
    &
        + 
        \frac{v_{[\mu } x_{\nu]}\,x_{[\alpha} y_{\beta]} }{(v\sdot x )^2(v\sdot y )}\,
        g_{\rho\sigma}\, \W^\sigma_{23}
        + 
        \frac{v_{[\mu } y_{\nu]}\,x_{[\alpha} y_{\beta]} }{(v\sdot x )(v\sdot y)^2}\,
        g_{\rho\sigma}\, \W^\sigma_{24}
        + 
        \frac{x_{[\mu } y_{\nu]}\,v_{[\alpha} x_{\beta]} }{(v\sdot x )^2(v\sdot y)}\,
        g_{\rho\sigma}\, \W^\sigma_{25}
        + 
        \frac{x_{[\mu } y_{\nu]}\,v_{[\alpha} y_{\beta]} }{(v\sdot x)(v\sdot y)^2}\,
        g_{\rho\sigma}\, \W^\sigma_{26}
    \nonumber\\*
    &
        + 
        \frac{x_{[\mu } y_{\nu]}\,x_{[\alpha} y_{\beta]} }{(v\sdot x)^2(v\sdot y)^2}\,
        g_{\rho\sigma}\, \W^\sigma_{27}
    \,.
    \label{eq:Geygen}
\end{align}
For convenience, we introduce the following notation:
\begin{align}
    \langle G_{\mu\nu} \Gamma \rangle
    &:=
    \frac{i}{m_B F_B(\mu) }
    \langle 0| \bar q(x)\, gG_{\mu\nu}(y)\,\Gamma\, h_v(0) |\bar B(v)\rangle \,,
    \\
    \C^\sigma_i 
    & :=
    v^\sigma \,\overline{\C}_i
    +
    \frac{x^\sigma}{v\sdot x}\, \widetilde{\C}_i
    +
    \frac{y^\sigma}{v\sdot y}\, \widehat{\C}_i \,,
\end{align}
with analogous definitions for $\X^\sigma_i$, $\Y^\sigma_i$, and $\W^\sigma_i$.
Each generalised DLCDA is understood to be a function of $v\sdot x$, $v\sdot y$,  and $x\sdot y$, e.g. \hbox{$\widetilde{\C}_1 := \widetilde{\C}_1( v\sdot x,v\sdot y, x\sdot y)$}.
Using gamma-matrix identities together with $P_+ h_v = h_v$, one readily obtains~\cite{Geyer:2005fb}
 \begin{align}
 \langle G_{\mu\nu}  \gamma_\rho i \sigma_{\alpha\beta} \rangle &=    
({i}/{2})\, \epsilon_{\alpha\beta \sigma\tau} 
\langle G_{\mu\nu} \gamma_5 \gamma_\rho i \sigma^{\sigma\tau} \rangle
,\\
\langle G_{\mu\nu} \gamma_5 \gamma_\alpha \rangle   &=   ({i}/{6})\,\epsilon_{\alpha\rho\sigma\tau}
 \langle G_{\mu\nu}  \gamma^\rho i \sigma^{\sigma\tau} \rangle
,\\
 \langle G_{\mu\nu} \gamma_5 \rangle &= v^\alpha \langle G_{\mu\nu} \gamma_5 \gamma_\alpha \rangle,\\
\langle G_{\mu\nu} \gamma_5 i \sigma_{\alpha\beta} \rangle 
  &=   v^\rho \langle G_{\mu\nu} \gamma_5 \gamma_\rho i \sigma_{\alpha\beta} 
\rangle
 +  2 \left(v_\alpha \langle G_{\mu\nu} \gamma_5 \gamma_\beta \rangle
       - v_\beta \langle G_{\mu\nu} \gamma_5 \gamma_\alpha \rangle \right),\\
\langle G_{\mu\nu} \gamma_\alpha \rangle
  &=   ({i}/{6})\,\epsilon_{\alpha\rho\sigma\tau}
 \langle G_{\mu\nu} \gamma_5 \gamma^\rho i \sigma^{\sigma\tau} \rangle,\\
\langle G_{\mu\nu} \rangle
  &=   v^\alpha \langle G_{\mu\nu} \gamma_\alpha \rangle,\\
\langle G_{\mu\nu} i \sigma_{\alpha\beta}\rangle
  &=   v^\rho \langle G_{\mu\nu} \gamma_\rho i \sigma_{\alpha\beta} \rangle
 + 2 \left(v_\alpha \langle G_{\mu\nu} \gamma_\beta \rangle
       - v_\beta \langle G_{\mu\nu}  \gamma_\alpha \rangle \right).
\end{align}
These relations motivate the choice of parametrising the matrix element $\langle G_{\mu\nu} \gamma_5 \gamma_\rho i \sigma_{\alpha\beta} \rangle$, since it can be related to matrix elements involving the complete basis of gamma matrices
\begin{align}
    \left\{ 
        \mathbf 1,\ \gamma^\mu,\ \sigma^{\mu\nu},\ \gamma^\mu\gamma^5,\ \gamma^5 
    \right\} .
\end{align}
In addition, imposing the Chisholm identity yields
\begin{align}
    \langle G_{\mu\nu}\gamma_5\gamma_\rho i\sigma_{\alpha\beta}\rangle
    =
    g_{\beta\rho}\langle G_{\mu\nu}\gamma_5\gamma_\alpha \rangle
    - 
    g_{\alpha\rho}\langle G_{\mu\nu}\gamma_5\gamma_\beta \rangle
    -i 
    \epsilon_{\rho\alpha\beta\sigma}\langle G_{\mu\nu}\gamma^\sigma\rangle\,.
    \label{eq:Chisholm}
\end{align}
By enforcing this constraint we find that the following generalised DLCDAs must vanish:
\begin{equation}
\begin{aligned}
\overline{\C}_{6} &= 0,\\
\widetilde{\X}_{7} &= 0, &
\widetilde{\X}_{9} &= 0, &
\widetilde{\X}_{10} &= 0,\\
\overline{\X}_{8} &= 0, &
\overline{\X}_{10} &= 0,\\
\widehat{\Y}_{13} &= 0, &
\widehat{\Y}_{15} &= 0, &
\widehat{\Y}_{16} &= 0,\\
\overline{\Y}_{14} &= 0, &
\overline{\Y}_{16} &= 0,\\
\widetilde{\W}_{20} &= 0, &
\widetilde{\W}_{22} &= 0, &
\widetilde{\W}_{23} &= 0, &
\widetilde{\W}_{24} &= 0, \\
\widetilde{\W}_{25} &= 0, &
\widetilde{\W}_{27} &= 0,\\
\widehat{\W}_{19} &= 0, &
\widehat{\W}_{21} &= 0, &
\widehat{\W}_{23} &= 0,\\
\widehat{\W}_{24} &= 0, &
\widehat{\W}_{26} &= 0, &
\widehat{\W}_{27} &= 0,\\
\overline{\W}_{21} &= 0, &\qquad
\overline{\W}_{22} &= 0, &\qquad
\overline{\W}_{25} &= 0, &\qquad
\overline{\W}_{26} &= 0.\\
\end{aligned}
\end{equation}
We also find the following relations:
\begin{equation}
\begin{aligned}
\label{eq:genset1}
\widetilde{\C}_{1} &= \widetilde{\C}_{3}, &
\widehat{\C}_{1} &= \widehat{\C}_{3}, &
\overline{\C}_{1} &= \overline{\C}_{3}, \\
\overline{\X}_{5} &= \widetilde{\C}_{6} = -\overline{\X}_{7}, &
\widetilde{\X}_{5} &= \widetilde{\X}_{8} = -\overline{\X}_{9}, &
\widehat{\X}_{5} &= \widetilde{\Y}_{14} = -\overline{\W}_{20}, \\
\overline{\Y}_{12} &= \widehat{\C}_{6} = -\overline{\Y}_{13}, &
\widetilde{\Y}_{12} &= \widehat{\X}_{8} = -\overline{\W}_{19}, &
\widehat{\Y}_{12} &= \widehat{\Y}_{14} = -\overline{\Y}_{15}, \\
\overline{\W}_{18} &= -\widetilde{\Y}_{13} = \widehat{\X}_{7}, &
\widetilde{\W}_{18} &= \widehat{\X}_{9} = -\widetilde{\W}_{19} , &
\widehat{\W}_{18} &= -\widetilde{\Y}_{15} = \widehat{\W}_{20}, \\
\overline{\W}_{23} &= -\widehat{\X}_{10} = \widetilde{\W}_{21}, &\qquad
\overline{\W}_{24} &= -\widetilde{\Y}_{16} = \widehat{\W}_{22}, &\qquad
\overline{\W}_{27} &= -\widetilde{\W}_{26} = \widehat{\W}_{25}.
\end{aligned}
\end{equation}
The following generalised DLCDAs are not constrained by \refeq{Chisholm}:
\begin{equation}
\begin{aligned}
\label{eq:genset2}
&
\overline{\C}_{2},
&&\widetilde{\C}_{2},
&&\widehat{\C}_{2},\\*
&
\overline{\X}_{4},
&&\widetilde{\X}_{4},
&&\widehat{\X}_{4},\\*
&
\overline{\Y}_{11},
&&\widetilde{\Y}_{11},
&&\widehat{\Y}_{11},\\*
&
\overline{\W}_{17},\qquad
&&\widetilde{\W}_{17},\qquad
&&\widehat{\W}_{17}.
\end{aligned}
\end{equation}
These 12 functions, together with the 15 functions appearing on the left-hand side of \refeq{genset1}, constitute a set of 27 generalised DLCDAs.
However, only 24 of combinations of these functions are genuinely independent.
This follows from the Schouten identity
\begin{align}
    g_{\eta[\mu}\epsilon_{\nu\rho\alpha\beta]}=0
\end{align}
which, when combined with the Dirac algebra and the HQET projector $P_+$, implies the presence of three independent residual redefinition freedoms in the set of 27 functions.
One can fix this redundancy, for example, by a suitable redefinition of the generalised DLCDAs such that $\overline \W_{23}=\overline \W_{24}=\overline \W_{27}=0$.
We do not provide the explicit form of these redefinitions here, as they would substantially complicate the relations presented in this appendix.
%
%
This redundancy must be taken into account when using the generalised DLCDAs in phenomenological applications, so that redundant functions are not treated as independent hadronic inputs.

In analogy with \refsec{defDDAs}, we first rewrite the parametrisation of \refeq{genTME} in a form suitable for practical calculations.
We then verify a posteriori that this parametrisation is consistent by matching it to \refeq{Geygen}:
\begin{align}
    \langle G_{\mu\nu} \Gamma \rangle
    & =
    \frac{i}{2} 
    \Tr\biggl\{\gamma_5 \Gamma P_+
    \biggl[
        \bigl( v_\mu \gamma_\nu - v_\nu \gamma_\mu \bigr)\bigl[ \Psi_1 - \Psi_2 \bigr]
        - i\sigma_{\mu\nu}\, \Psi_2
        \nonumber\\
        &
        -
        \frac{x_\mu v_\nu - x_\nu v_\mu}{v\sdot x}\,X_1
        +
        \frac{x_\mu \gamma_\nu - x_\nu \gamma_\mu}{v\sdot x}\,\big[W_1+Y_1\big]
        -
        \frac{i\,\epsilon_{\mu\nu\alpha\beta}\,x^\alpha v^\beta\gamma_5}{v\sdot x}\,\widetilde{X}_1
        +
        \frac{i\,\epsilon_{\mu\nu\alpha\beta}\,x^\alpha \gamma^\beta \gamma_5}{v\sdot x}\,\widetilde{Y}_1
        \nonumber\\
        &
        -
        \frac{(x_\mu v_\nu - x_\nu v_\mu)\slashed{x}}{(v\sdot x)^2}\,W_1
        +
        \frac{(x_\mu \gamma_\nu - x_\nu \gamma_\mu)\slashed{x}}{(v\sdot x)^2}\,Z_1
        \nonumber\\
        &
        -
        \frac{y_\mu v_\nu - y_\nu v_\mu}{v\sdot y}\,X_2
        +
        \frac{y_\mu \gamma_\nu - y_\nu \gamma_\mu}{v\sdot y}\,\big[W_2+Y_2\big]
        -
        \frac{i\,\epsilon_{\mu\nu\alpha\beta}\,y^\alpha v^\beta\gamma_5}{v\sdot y}\,\widetilde{X}_2
        +
        \frac{i\,\epsilon_{\mu\nu\alpha\beta}\,y^\alpha \gamma^\beta \gamma_5}{v\sdot y}\,\widetilde{Y}_2
        \nonumber\\
        &
        -
        \frac{(y_\mu v_\nu - y_\nu v_\mu)\slashed{y}}{(v\sdot y)^2}\,W_2
        +
        \frac{(y_\mu \gamma_\nu - y_\nu \gamma_\mu)\slashed{y}}{(v\sdot y)^2}\,Z_2
        \nonumber\\
        &
        -
        \frac{x_\mu y_\nu - x_\nu y_\mu}{(v\sdot x)(v\sdot y)}\,X_3
        -
        \frac{i\,\epsilon_{\mu\nu\alpha\beta}\,x^\alpha y^\beta\gamma_5}{(v\sdot x)(v\sdot y)}\,\widetilde{X}_3
        \nonumber\\
        &
        -
        \frac{i\,\epsilon_{\mu\nu\alpha\beta}\,x^\alpha y^\beta\,\slashed{x}\,\gamma_5}
        {(x\sdot y) (v\sdot x)}\,
        \widetilde{W}_1
        -
        \frac{i\,\epsilon_{\mu\nu\alpha\beta}\,x^\alpha y^\beta\,\slashed{y}\,\gamma_5}
        {(x\sdot y)(v\sdot y)}\,
        \widetilde{W}_2
        \nonumber\\
        &
        -
        \frac{i\,\epsilon_{\alpha\beta\lambda\tau}\,x^\alpha y^\beta v^\lambda \gamma^\tau\,\gamma_5}{(v\sdot x)(v\sdot y)}\,
        \left(
            \frac{x_\mu v_\nu - x_\nu v_\mu}{(v\sdot x)}
            \widetilde{W}_3
            +
            \frac{y_\mu v_\nu - y_\nu v_\mu}{(v\sdot y)}
            \widetilde{W}_4
            +
            \frac{x_\mu y_\nu - x_\nu y_\mu}{(v\sdot x) (v\sdot y)}
            \widetilde{W}_5
        \right)
        \nonumber\\
        &
        -
        \frac{(x_\mu v_\nu - x_\nu v_\mu)\slashed{y}}{(v\sdot x)(v\sdot y)}\,W_3
        -
        \frac{(y_\mu v_\nu - y_\nu v_\mu)\slashed{x}}{(v\sdot x)(v\sdot y)}\,W_4
        -
        \frac{(x_\mu y_\nu - x_\nu y_\mu)\,\slashed{x}}{(v\sdot x)^2 (v\sdot y)}\,W_5
        -
        \frac{(x_\mu y_\nu - x_\nu y_\mu)\,\slashed{y}}{(v\sdot x)(v\sdot y)^2}\,W_6
        \nonumber\\
        &
        +
        \frac{(x_\mu \gamma_\nu - x_\nu \gamma_\mu)\slashed{y}}{(v\sdot x)(v\sdot y)}\,Z_3
        +
        \frac{(y_\mu \gamma_\nu - y_\nu \gamma_\mu)\slashed{x}}{(v\sdot x)(v\sdot y)}\,Z_4
        \,
        \nonumber\\
    &
    \biggr]\biggr\}(v\sdot x,v\sdot y, x\sdot y;\mu)\,.
    \label{eq:Brgen}
\end{align}
We verify that the 27 generalised DLCDAs appearing in the equation above are independent by relating them to the parametrisation introduced in \refeq{Geygen}.
This statement holds up to the three redefinition freedoms implied by the Chisholm identity discussed above.
In this case, these redefinitions can be used to effectively remove $\widetilde{W}_3$, $\widetilde{W}_4$, and $\widetilde{W}_5$ from \refeq{Brgen}
The relation between the two parametrisations can be established by computing the matrix elements
\begin{align}
    \langle G_{\mu\nu} \gamma_5 \rangle \,, \qquad
    \langle G_{\mu\nu} \gamma_5 \gamma_\alpha \rangle \,, \qquad
    \langle G_{\mu\nu} \gamma_5 \sigma_{\alpha\beta} \rangle \,, \qquad
    \langle G_{\mu\nu} \gamma_5 \gamma_\rho \sigma_{\alpha\beta}  \rangle \,,
\end{align}
using \refeqa{Geygen}{Brgen}, and subsequently solving the resulting system of equations.
We obtain
\begin{align*}
\Psi_1 &= \overline{\C}_2 ,
&\qquad
\Psi_2 &= \overline{\C}_1-\overline{\X}_5-\overline{\Y}_{12} ,
&\qquad
X_1 &= -\overline{\X}_4 ,\\[1mm]
W_1 &= \widetilde{\X}_5-\widetilde{\X}_4 ,
&\qquad
Y_1 &= \widetilde{\C}_2+\widetilde{\X}_4+\widetilde{\Y}_{12} ,
&\qquad
\widetilde{X}_1 &= -\overline{\X}_5 ,\\[1mm]
\widetilde{Y}_1 &= -\widetilde{\C}_1+\frac{\widetilde{\W}_{18}\,x\sdot y}{v\sdot x\,v\sdot y}+\overline{\W}_{18}-\overline{\X}_5 ,
&\qquad
Z_1 &= \widetilde{\X}_5 ,
&\qquad
X_2 &= -\overline{\Y}_{11} ,\\[1mm]
W_2 &= \widehat{\Y}_{12}-\widehat{\Y}_{11} ,
&\qquad
Y_2 &= \widehat{\C}_2+\widehat{\X}_5+\widehat{\Y}_{11} ,
&\qquad
\widetilde{X}_2 &= -\overline{\Y}_{12} ,\\[1mm]
\widetilde{Y}_2 &= -\widehat{\C}_1-\frac{\widehat{\W}_{18}\,x\sdot y}{v\sdot x\,v\sdot y}-\overline{\W}_{18}-\overline{\Y}_{12} ,
&\qquad
Z_2 &= \widehat{\Y}_{12} ,
&\qquad
X_3 &= \overline{\W}_{17}-\widehat{\X}_5+\widetilde{\Y}_{12} ,\\[1mm]
\widetilde{X}_3 &= \overline{\W}_{18} ,
&\qquad
W_3 &= \widetilde{\Y}_{12}-\widehat{\X}_4 ,
&\qquad
W_4 &= \widehat{\X}_5-\widetilde{\Y}_{11} ,\\[1mm]
W_5 &= \widetilde{\W}_{17} ,
&\qquad
W_6 &= \widehat{\W}_{17} ,
&\qquad
\widetilde{W}_1 &= \frac{\widetilde{\W}_{18}\,x\sdot y}{v\sdot x\,v\sdot y} ,\\[1mm]
\widetilde{W}_2 &= \frac{\widehat{\W}_{18}\,x\sdot y}{v\sdot x\,v\sdot y} ,
&\qquad
\widetilde{W}_3 &= -\overline{\W}_{23} ,
&\qquad
\widetilde{W}_4 &= -\overline{\W}_{24} ,\\[1mm]
\widetilde{W}_5 &= \overline{\W}_{27} ,
&\qquad
Z_3 &= \widetilde{\Y}_{12} ,
&\qquad
Z_4 &= \widehat{\X}_5 .
\end{align*}
This system of equations can also be inverted:
\begin{align*}
\overline{\C}_1 &= \Psi_2-\widetilde{X}_1-\widetilde{X}_2 ,
&\qquad
\widetilde{\C}_1 &= \widetilde{W}_1+\widetilde{X}_1+\widetilde{X}_3-\widetilde{Y}_1 ,
&\qquad
\widehat{\C}_1 &= -\widetilde{W}_2+\widetilde{X}_2-\widetilde{X}_3-\widetilde{Y}_2 ,\\[1mm]
\overline{\X}_5 &= -\widetilde{X}_1 ,
&\qquad
\widetilde{\X}_5 &= Z_1 ,
&\qquad
\widehat{\X}_5 &= Z_4 ,\\[1mm]
\overline{\Y}_{12} &= -\widetilde{X}_2 ,
&\qquad
\widetilde{\Y}_{12} &= Z_3 ,
&\qquad
\widehat{\Y}_{12} &= Z_2 ,\\[1mm]
\overline{\W}_{18} &= \widetilde{X}_3 ,
&\qquad
\widetilde{\W}_{18} &= \frac{\widetilde{W}_1\,v\sdot x\,v\sdot y}{x\sdot y} ,
&\qquad
\widehat{\W}_{18} &= \frac{\widetilde{W}_2\,v\sdot x\,v\sdot y}{x\sdot y} ,\\[1mm]
\overline{\W}_{23} &= -\widetilde{W}_3 ,
&\qquad
\overline{\W}_{24} &= -\widetilde{W}_4 ,
&\qquad
\overline{\W}_{27} &= \widetilde{W}_5 ,\\[1mm]
\overline{\C}_2 &= \Psi_1 ,
&\qquad
\widehat{\C}_2 &= W_2+Y_2-Z_2-Z_4 ,
&\qquad
\overline\X_4 &= -X_1 ,\\[1mm]
\widehat{\X}_4 &= Z_3-W_3 ,
&\qquad
\widetilde{\X}_4 &= Z_1-W_1 ,
&\qquad
\widetilde{\C}_2 &= W_1+Y_1-Z_1-Z_3 ,\\[1mm]
\overline{\W}_{17} &= X_3-Z_3+Z_4 ,
&\qquad
\widehat{\W}_{17} &= W_6 ,
&\qquad
\widetilde{\W}_{17} &= W_5 ,\\[1mm]
\overline{\Y}_{11} &= -X_2 ,
&\qquad
\widehat{\Y}_{11} &= Z_2-W_2 ,
&\qquad
\widetilde{\Y}_{11} &= Z_4-W_4 .
\end{align*}
This system is algebraically invertible and therefore establishes a one-to-one correspondence between the two 27-function sets in \refeqs{genset1}{genset2} and \refeq{Brgen}.
We note that all formulas derived in this appendix up to this point hold for arbitrary four-vectors $x$ and $y$, i.e.\ they are not restricted to the light-cone.
The only exceptions are the expressions for $\widetilde{\C}_1$, $\widehat{\C}_1$, $\widetilde{Y}_1$, and $\widetilde{Y}_2$, which in the general case should read
\begin{align*}
    \widetilde{\C}_1 &= \widetilde{W}_1+\frac{\widetilde{W}_2\,y^2\,v\sdot x}{v\sdot y \, x\sdot y}+\widetilde{X}_1+\widetilde{X}_3-\widetilde{Y}_1 ,
    &
    \widehat{\C}_1 &=-\frac{\widetilde{W}_1\,x^2\,v\sdot y}{v\sdot x \, x\sdot y} -\widetilde{W}_2+\widetilde{X}_2-\widetilde{X}_3-\widetilde{Y}_2 
    \\
    \widetilde{Y}_1 &= -\widetilde{\C}_1+\frac{\widetilde{\W}_{18}\,x\sdot y}{v\sdot x\,v\sdot y}+\overline{\W}_{18}+\frac{\widehat{\W}_{18}\,y^2}{(v\sdot y)^2}-\overline{\X}_5 ,
    &
    \widetilde{Y}_2 &= -\widehat{\C}_1-\frac{\widehat{\W}_{18}\,x\sdot y}{v\sdot x\,v\sdot y}-\overline{\W}_{18}-\frac{\widetilde{\W}_{18}\,x^2}{(v\sdot x)^2}-\overline{\Y}_{12} .
\end{align*}
In this more general setting, the corresponding functions are to be understood as depending on $x^2$, $y^2$, $v\sdot x$, $v\sdot y$, and $x\sdot y$.
For example, \hbox{$\widetilde{\C}_1 := \widetilde{\C}_1(x^2, y^2, v\sdot x,v\sdot y, x\sdot y)$}.
\\

In the limit $x_\mu \sim n_\mu \sim y_\mu$, the generalised parametrisation of \refeq{Brgen} must reduce to the standard one in Eq.~(2.7) of Ref.~\cite{Braun:2017liq}.
Matching the Lorentz and Dirac structures gives
\begin{align}
\begin{aligned}
    \Psi_A &= \Psi_1 \,,&
    \Psi_V &= \Psi_2 \,,\\
    X_A &= X_1 + X_2\,,&
    Y_A &= Y_1 + Y_2 - (W_3 + W_4)\,,\\
    \widetilde{X}_A &= \widetilde{X}_1 + \widetilde{X}_2 \,,&
    \widetilde{Y}_A &= \widetilde{Y}_1 + \widetilde{Y}_2 \,,\\
    W &= W_1 + W_2 + W_3 + W_4 \,,&
    Z &= Z_1 + Z_2 + Z_3 + Z_4 \,.
\end{aligned}
\end{align}
The functions multiplying Lorentz structures that vanish when $x_\mu$ and $y_\mu$ are collinear remain unconstrained:
\begin{align}
    X_3,\quad
    \widetilde{X}_3,\quad
    \widetilde{W}_1,\quad
    \widetilde{W}_2,\quad
    \widetilde{W}_3,\quad
    \widetilde{W}_4,\quad
    \widetilde{W}_5,\quad
    W_5,\quad
    W_6\,.
\end{align}
In the limit $x_\mu \sim n_\mu$ and $y_\mu \sim \bar{n}_\mu$ the distribution amplitudes of \refeq{Brnnbar} and \refeq{Brgen} can be related:
\begin{align}
Y_x &= \tfrac{1}{2}\Bigl(
  2W_1 + 2W_2 - 2W_3 - 4W_6
  - 2\widetilde{W}_1 + 2\widetilde{W}_2
  - X_1 + \Psi_1 + X_2 - 2X_3
  + 2Y_1 - 4Z_4
\Bigr) , \\
Y_y &= \tfrac{1}{2}\Bigl(
  2W_3 + 4W_6
  + X_1 + \Psi_1 - X_2 + 2X_3
  + 2Y_2 - 4Z_3
\Bigr) , \\
Z_x &= \tfrac{1}{2}\Bigl(
  2W_2 - 2W_3 - 4W_6
  - 2\widetilde{W}_1 + 2\widetilde{W}_2
  - X_1 + X_2 - 2X_3
  + 2Z_1 + 2Z_3 - 4Z_4
\Bigr) , \\
Z_y &= \tfrac{1}{2}\Bigl(
  -2W_2 + 2W_3 + 4W_6
  + X_1 - X_2 + 2X_3
  + 2Z_2 - 4Z_3 + 2Z_4
\Bigr) , \\
\widetilde{Y}_x &= \tfrac{1}{2}\Bigl(
  2W_2 - 2W_3 - 4W_6
  - 2\widetilde{W}_1 + 2\widetilde{W}_2
  - X_1 + X_2 - \Psi_2 - 2X_3
  + 2\widetilde{Y}_1 + 4Z_3 - 4Z_4
\Bigr) , \\
\widetilde{Y}_y &= \tfrac{1}{2}\Bigl(
  -2W_2 + 2W_3 + 4W_6
  + X_1 - X_2 - \Psi_2 + 2X_3
  + 2\widetilde{Y}_2 - 4Z_3 + 4Z_4
\Bigr) , \\
\widetilde{X}_{xy} &= \tfrac{1}{2}\Bigl(
  2W_2 - 2W_3 - 4W_6
  + 2\widetilde{W}_2
  - X_1 + X_2 - 2X_3
  + \widetilde{X}_1 - \widetilde{X}_2 + 2\widetilde{X}_3
  + 4Z_3 - 4Z_4
\Bigr) , \\
W_{xy} &= \tfrac{1}{2}\Bigl(
  W_1 + W_2 - W_3 - W_4
  + 2W_5 - 2W_6
  - \widetilde{W}_1 + \widetilde{W}_2
\Bigr) .
\end{align}

\appsection{First moments of the DLCDAs and EOM constraints}
\label{app:DA_first_moments}

As discussed in \refsec{models}, the parametrisation of the three-derivative matrix element in \refeq{three_derivative_parametrisation} allows one to derive the first moments in the momentum variables $\omega_1$ and $\bar{\omega}_2$ through \refeqa{w1_moments}{w2_moments}.
These moments are given by
\begin{align}
    \langle \omega_1 \rangle_{\Phi_3} &= 2 (f_2 - f_3) + g_2 - g_3 - 4 h_1 \,, \\
    \langle \omega_1 \rangle_{\Phi_4} &= 2 (f_2 - f_3) + g_2 - g_3 - 2 (h_2 - h_3) \,, \\
    \langle \omega_1 \rangle_{\Psi_4} &= -e_2 + e_3 + g_2 - g_3 - 2 h_1 - h_2 + h_3 \,, \\
    \langle \omega_1 \rangle_{\widetilde{\Psi}_4} &= -2 h_1 + 2 h_4 \,, \\
    \langle \omega_1 \rangle_{\Phi_5} &= g_2 - g_3 \,, \\
    \langle \omega_1 \rangle_{\Psi_5} &= -e_2 + e_3 - 2 f_2 + 2 f_3 - g_2 + g_3 + 2 h_1 + h_2 - h_3 \,, \\
    \langle \omega_1 \rangle_{\widetilde{\Psi}_5} &= 2 h_1 + 2 h_4 \,, \\
    \langle \omega_1 \rangle_{\Phi_6} &= g_2 - g_3 - 4 h_1 - 2 h_2 + 2 h_3 \,, \\
    \langle \bar{\omega}_2 \rangle_{\Phi_3} &= g_1 - 2 g_2 + g_3 + 6 h_1 - 6 h_3 \,, \\
    \langle \bar{\omega}_2 \rangle_{\Phi_4} &= g_1 - 2 g_2 + g_3 \,, \\
    \langle \bar{\omega}_2 \rangle_{\Psi_4} &= e_1 - 2 e_2 + e_3 + 2 f_1 - 4 f_2 + 2 f_3 + g_1 - 2 g_2 + g_3 + 3 h_1 - 3 h_3 \,, \\
    \langle \bar{\omega}_2 \rangle_{\widetilde{\Psi}_4} &= 2 (h_1 - h_3) \,, \\
    \langle \bar{\omega}_2 \rangle_{\Phi_5} &= 2 f_1 - 4 f_2 + 2 f_3 + g_1 - 2 g_2 + g_3 + 2 h_1 - 2 h_3 \,, \\
    \langle \bar{\omega}_2 \rangle_{\Psi_5} &= e_1 - 2 e_2 + e_3 - g_1 + 2 g_2 - g_3 - 3 h_1 + 3 h_3 \,, \\
    \langle \bar{\omega}_2 \rangle_{\widetilde{\Psi}_5} &= 2 h_3 - 2 h_1 \,, \\
    \langle \bar{\omega}_2 \rangle_{\Phi_6} &= 2 f_1 - 4 f_2 + 2 f_3 + g_1 - 2 g_2 + g_3 + 4 h_1 - 4 h_3 \,.
\end{align}
Additionally, we provide here the relations among the various parameters obtained from the EOM.
By applying the quark EOM, 8 of the 14 parameters can be eliminated and expressed in terms of the remaining quantities
$e_1, f_3, g_1, g_2, g_3, h_2, \bar{\Lambda}, \lambda_E^2, \lambda_H^2$:
\begin{align}
d   &= -2 (2 g_1 + g_2 + g_3) \,, \\
e_2 &= f_3 + g_2 + g_3 + 3 h_2 + \bar{\Lambda} \frac{2 \bar{\Lambda}^2 + 2 \lambda_E^2 - \lambda_H^2}{6} \,, \\
e_3 &= -f_3 - \bar{\Lambda} \frac{\bar{\Lambda}^2 + \lambda_E^2 + \lambda_H^2}{3} \,, \\
f_1 &= -g_1 + h_2 - \bar{\Lambda} \frac{2 \bar{\Lambda}^2 + \lambda_H^2}{6} \,, \\
f_2 &= -e_1 - f_3 + 4 (g_1 - h_2) + \frac{2}{3} \bar{\Lambda} (2 \bar{\Lambda}^2 + \lambda_H^2) \,, \\
h_1 &= -e_1 - f_3 + 4 g_1 + g_2 - 4 h_2 + \bar{\Lambda} \frac{10 \bar{\Lambda}^2 + 2 \lambda_E^2 + 3 \lambda_H^2}{6} \,, \\
h_3 &= -\frac{\bar{\Lambda} \lambda_H^2}{6} \,, \\
h_4 &= 2 e_1 + 3 f_3 - 8 g_1 - g_2 + 8 h_2 - \bar{\Lambda} \frac{18 \bar{\Lambda}^2 + 2 \lambda_E^2 + 7 \lambda_H^2}{6} \,.
\end{align}
Furthermore, by employing the gluon EOM and neglecting four-particle contributions, we obtain two further independent relations:
\begin{align}
    f_3 &= - e_1 + 2 e_2 - e_3 - f_1 + 2 f_2\,,\\
    g_3 &= 3 e_1 - 6 e_2 + 3 e_3 - g_1 + 2 g_2 - 3 h_1 + 3 h_3\,.
\end{align}

\appsection{Useful formulas for transformation between momentum and position space}
\label{app:laplace}

\begin{table}[h!]
\begin{center}
\renewcommand{\arraystretch}{2}
\begin{tabular}{|c|c|}
\hline 
$f(\omega)$  & ${\cal L}[f](s=iz)$
\\
\hline
$\frac{1}{\omega_0} \left( \frac{\omega}{\omega_0}\right)^n e^{-\omega/\omega_0}$ &
$\frac{\Gamma(1+n)}{\left(1+ s \omega_0 \right)^{n+1}}$
\\
$1$ & $\frac{1}{s} $
\\
$\ln \frac{\mu}{\omega} $ & $\frac{1}{s} \, \ln(\hat \mu s)$
\\
$\ln^2 \frac{\mu}{\omega}  $ & $ 
\frac{\ln^2(\hat\mu s) + \pi^2/6}{s} $
\\
\hline
\end{tabular}
\qquad 
\begin{tabular}{|c|c|}
\hline 
$f(\omega)$  & ${\cal L}[f](s=iz)$
\\
\hline 
$\frac{e^{-\omega/\hat\mu_F}}{\hat\mu_F}$ 
& 
$ \frac{1}{1+s\hat \mu_F} $
\\
$\frac{e^{-\omega/\hat\mu_F}}{\hat\mu_F} \, \ln \frac{\mu_F}{\omega} $ & $\frac{\ln(1+\hat\mu_F s)}{1+\hat\mu_F s} $
\\
$\frac{e^{-\omega/\hat\mu_F}}{\hat\mu_F} \, \ln^2 \frac{\mu_F}{\omega} $
&
$
\frac{\ln^2(1+\hat\mu_F s) + \pi^2/6}{1+\hat\mu_F s} $
\\
$\frac{e^{-\omega/\hat \mu_F}-1}{\omega}$
& 
$ 
\ln \frac{\hat \mu_F s}{1+\hat \mu_F s}
$
\\
\hline 
\end{tabular}
\end{center} 
\caption{Laplace transform of various functions appearing in the modelling of the radiative tail
for the DLCDA $\Phi_3(\omega_1,\bar{\omega}_2;\mu)$.
Here we use again the shorthand notation $\hat \mu := \mu e^{\gamma_E}$.
\label{tab:laplace}
}
\end{table}

In this appendix we collect some useful formulas that
allow us to translate our results for the modelling of the radiative tail from momentum to position space and vice versa. As we restrict ourselves to positive values of $\omega_1$ and $\bar{\omega}_2$, this amounts to 
performing (inverse) Laplace transforms from 
a momentum variable $\omega$ 
to an imaginary position $s=iz$.
The relevant expressions are collected in \reftab{laplace}.

\bibliographystyle{JHEP}
\bibliography{refs}

@article{Feldmann:2025dcs,
    author = "Feldmann, Thorsten and Vladimirov, Daniel",
    title = "{Radiative tail of the three-particle light-cone distribution amplitudes for the $\Lambda_b$ baryon in HQET}",
    eprint = "2505.02570",
    archivePrefix = "arXiv",
    primaryClass = "hep-ph",
    reportNumber = "SI-HEP-2025-026, P3H-25-022",
    doi = "10.1007/JHEP07(2025)108",
    journal = "JHEP",
    volume = "07",
    pages = "108",
    year = "2025"
}

@article{Feldmann:2023plv,
    author = "Feldmann, Thorsten and Gubernari, Nico",
    title = "{Non-factorisable contributions of strong-penguin operators in $\Lambda_b \to \Lambda \ell^+\ell^-$ decays}",
    eprint = "2312.14146",
    archivePrefix = "arXiv",
    primaryClass = "hep-ph",
    doi = "10.1007/JHEP03(2024)152",
    journal = "JHEP",
    volume = "03",
    pages = "152",
    year = "2024"
}

@article{Bell:2013tfa,
    author = "Bell, Guido and Feldmann, Thorsten and Wang, Yu-Ming and Yip, Matthew W Y",
    title = "{Light-Cone Distribution Amplitudes for Heavy-Quark Hadrons}",
    eprint = "1308.6114",
    archivePrefix = "arXiv",
    primaryClass = "hep-ph",
    reportNumber = "OUTP-13-15P, SI-HEP-2013-05, QFET-2013-05, TTK-13-19, SFB-CPP-13-60, IPPP-13-55, DCPT-13-110, TUM-HEP-899-13",
    doi = "10.1007/JHEP11(2013)191",
    journal = "JHEP",
    volume = "11",
    pages = "191",
    year = "2013"
}

@article{Braun:2017liq,
    author = "Braun, V. M. and Ji, Yao and Manashov, A. N.",
    title = "{Higher-twist $B$-meson Distribution Amplitudes in HQET}",
    eprint = "1703.02446",
    archivePrefix = "arXiv",
    primaryClass = "hep-ph",
    reportNumber = "DESY-17-037",
    doi = "10.1007/JHEP05(2017)022",
    journal = "JHEP",
    volume = "05",
    pages = "022",
    year = "2017"
}

@article{Qin:2022rlk,
    author = "Qin, Qin and Shen, Yue-Long and Wang, Chao and Wang, Yu-Ming",
    title = "{Deciphering the Long-Distance Penguin Contribution to $B_{d,s}\to\gamma\gamma$ Decays}",
    eprint = "2207.02691",
    archivePrefix = "arXiv",
    primaryClass = "hep-ph",
    doi = "10.1103/PhysRevLett.131.091902",
    journal = "Phys. Rev. Lett.",
    volume = "131",
    number = "9",
    pages = "091902",
    year = "2023"
}

@article{Bartocci:2024bbf,
    author = {Bartocci, Riccardo and B{\"o}er, Philipp and Hurth, Tobias},
    title = "{Renormalisation group evolution of the shape function g$_{17}$ in $ \overline{B}\to {X}_s\gamma $ and $ \overline{B}\to {X}_s{\ell}^{+}{\ell}^{-} $ at subleading power}",
    eprint = "2411.16634",
    archivePrefix = "arXiv",
    primaryClass = "hep-ph",
    reportNumber = "MITP/24-078, CERN-TH-2024-191",
    doi = "10.1007/JHEP04(2025)066",
    journal = "JHEP",
    volume = "04",
    pages = "066",
    year = "2025"
}

@article{Geyer:2005fb,
    author = "Geyer, Bodo and Witzel, Oliver",
    title = "{$B$-meson distribution amplitudes of geometric twist vs. dynamical twist}",
    eprint = "hep-ph/0502239",
    archivePrefix = "arXiv",
    doi = "10.1103/PhysRevD.72.034023",
    journal = "Phys. Rev. D",
    volume = "72",
    pages = "034023",
    year = "2005"
}

@article{Kawamura:2001jm,
    author = "Kawamura, Hiroyuki and Kodaira, Jiro and Qiao, Cong-Feng and Tanaka, Kazuhiro",
    title = "{$B$-meson light cone distribution amplitudes in the heavy quark limit}",
    eprint = "hep-ph/0109181",
    archivePrefix = "arXiv",
    reportNumber = "DESY-01-135, HUPD-0108, JUPD-0191",
    doi = "10.1016/S0370-2693(01)01299-0",
    journal = "Phys. Lett. B",
    volume = "523",
    pages = "111",
    year = "2001",
    note = "[Erratum: Phys.Lett.B 536, 344--344 (2002)]"
}

@article{Braun:1989iv,
    author = "Braun, Vladimir M. and Filyanov, I. E.",
    title = "{Conformal Invariance and Pion Wave Functions of Nonleading Twist}",
    reportNumber = "LENINGRAD-89-1567",
    doi = "10.1007/BF01554472",
    journal = "Z. Phys. C",
    volume = "48",
    pages = "239--248",
    year = "1990"
}

@article{Grozin:1996pq,
    author = "Grozin, A. G. and Neubert, M.",
    title = "{Asymptotics of heavy meson form-factors}",
    eprint = "hep-ph/9607366",
    archivePrefix = "arXiv",
    reportNumber = "BUDKER-INP-1996-45, BINP-96-45, CERN-TH-96-144",
    doi = "10.1103/PhysRevD.55.272",
    journal = "Phys. Rev. D",
    volume = "55",
    pages = "272--290",
    year = "1997"
}

@article{Nishikawa:2011qk,
    author = "Nishikawa, Tetsuo and Tanaka, Kazuhiro",
    title = "{QCD Sum Rules for Quark-Gluon Three-Body Components in the $B$ Meson}",
    eprint = "1109.6786",
    archivePrefix = "arXiv",
    primaryClass = "hep-ph",
    reportNumber = "J-PARC-TH-0032",
    doi = "10.1016/j.nuclphysb.2013.12.007",
    journal = "Nucl. Phys. B",
    volume = "879",
    pages = "110--142",
    year = "2014"
}

@article{Huang:2023jdu,
    author = "Huang, Yong-Kang and Ji, Yao and Shen, Yue-Long and Wang, Chao and Wang, Yu-Ming and Zhao, Xue-Chen",
    title = "{Renormalization-Group Evolution for the Three-Particle $B$-Meson Soft Function}",
    eprint = "2312.15439",
    archivePrefix = "arXiv",
    primaryClass = "hep-ph",
    reportNumber = "TUM-HEP-1489/23",
    doi = "10.1103/PhysRevLett.133.171901",
    journal = "Phys. Rev. Lett.",
    volume = "133",
    number = "17",
    pages = "171901",
    year = "2024"
}

@article{BenekeBoeer:2026,
    author = "Beneke, Martin and Böer, Philipp",
    note = "(work in progress)",
    journal = ", Work in progress$\!\!$"
}

@article{Lange:2003ff,
    author = "Lange, Bjorn O. and Neubert, Matthias",
    title = "{Renormalization group evolution of the B meson light cone distribution amplitude}",
    eprint = "hep-ph/0303082",
    archivePrefix = "arXiv",
    reportNumber = "CLNS-03-1822",
    doi = "10.1103/PhysRevLett.91.102001",
    journal = "Phys. Rev. Lett.",
    volume = "91",
    pages = "102001",
    year = "2003"
}

@article{Rahimi:2020zzo,
    author = "Rahimi, Muslem and Wald, Marcel",
    title = "{QCD sum rules for parameters of the $B$-meson distribution amplitudes}",
    eprint = "2012.12165",
    archivePrefix = "arXiv",
    primaryClass = "hep-ph",
    reportNumber = "P3H-20-082, SI-HEP-2020-35",
    doi = "10.1103/PhysRevD.104.016027",
    journal = "Phys. Rev. D",
    volume = "104",
    number = "1",
    pages = "016027",
    year = "2021"
}

@article{Beneke:2022msp,
    author = {Beneke, Martin and B{\"o}er, Philipp and Toelstede, Jan-Niklas and Vos, Keri K.},
    title = "{Light-cone distribution amplitudes of heavy mesons with QED effects}",
    eprint = "2204.09091",
    archivePrefix = "arXiv",
    primaryClass = "hep-ph",
    reportNumber = "TUM-HEP-1397/22, MITP-21-062, Nikhef-2022-004",
    doi = "10.1007/JHEP08(2022)020",
    journal = "JHEP",
    volume = "08",
    pages = "020",
    year = "2022"
}

@article{Khodjamirian:2023wol,
    author = "Khodjamirian, Alexander and Meli{\'c}, Bla{\v{z}}enka and Wang, Yu-Ming",
    title = "{A guide to the QCD light-cone sum rules for $b$-quark decays}",
    eprint = "2311.08700",
    archivePrefix = "arXiv",
    primaryClass = "hep-ph",
    reportNumber = "SI-HEP-2023-25, P3H-23-087; RBI-ThPhys-2023-37",
    doi = "10.1140/epjs/s11734-023-01046-6",
    journal = "Eur. Phys. J. ST",
    volume = "233",
    number = "2",
    pages = "271--298",
    year = "2024"
}

@article{Colangelo:2000dp,
    author = "Colangelo, Pietro and Khodjamirian, Alexander",
    editor = "Shifman, M. and Ioffe, Boris",
    title = "{QCD sum rules, a modern perspective}",
    eprint = "hep-ph/0010175",
    archivePrefix = "arXiv",
    reportNumber = "CERN-TH-2000-296, BARI-TH-2000-394",
    doi = "10.1142/9789812810458_0033",
    pages = "1495--1576",
    month = "10",
    year = "2000"
}

@article{Piscopo:2023opf,
    author = "Piscopo, Maria Laura and Rusov, Aleksey V.",
    title = "{Non-factorisable effects in the decays $ {\overline{B}}_s^0\to {D}_s^{+}{\pi}^{-} $ and $ {\overline{B}}^0\to {D}^{+}{K}^{-} $ from LCSR}",
    eprint = "2307.07594",
    archivePrefix = "arXiv",
    primaryClass = "hep-ph",
    reportNumber = "SI-HEP-2023-15",
    doi = "10.1007/JHEP10(2023)180",
    journal = "JHEP",
    volume = "10",
    pages = "180",
    year = "2023"
}

@article{Beneke:1999br,
    author = "Beneke, M. and Buchalla, G. and Neubert, M. and Sachrajda, Christopher T.",
    title = "{QCD factorization for $B \to \pi \pi$ decays: Strong phases and CP violation in the heavy quark limit}",
    eprint = "hep-ph/9905312",
    archivePrefix = "arXiv",
    reportNumber = "SLAC-PUB-8146, CERN-TH-99-126, SHEP-99-04",
    doi = "10.1103/PhysRevLett.83.1914",
    journal = "Phys. Rev. Lett.",
    volume = "83",
    pages = "1914--1917",
    year = "1999"
}

@article{Beneke:2000ry,
    author = "Beneke, M. and Buchalla, G. and Neubert, M. and Sachrajda, Christopher T.",
    title = "{QCD factorization for exclusive, nonleptonic B meson decays: General arguments and the case of heavy light final states}",
    eprint = "hep-ph/0006124",
    archivePrefix = "arXiv",
    reportNumber = "CERN-TH-2000-159, CLNS-00-1675, PITHA-00-06, SHEP-00-06",
    doi = "10.1016/S0550-3213(00)00559-9",
    journal = "Nucl. Phys. B",
    volume = "591",
    pages = "313--418",
    year = "2000"
}

@article{Beneke:2001ev,
    author = "Beneke, M. and Buchalla, G. and Neubert, M. and Sachrajda, Christopher T.",
    title = "{QCD factorization in $B \to \pi K, \pi \pi$ decays and extraction of Wolfenstein parameters}",
    eprint = "hep-ph/0104110",
    archivePrefix = "arXiv",
    reportNumber = "CERN-TH-2001-107, CLNS-01-1728, PITHA-01-01, SHEP-01-11",
    doi = "10.1016/S0550-3213(01)00251-6",
    journal = "Nucl. Phys. B",
    volume = "606",
    pages = "245--321",
    year = "2001"
}

@article{Descotes-Genon:2002crx,
    author = "Descotes-Genon, S. and Sachrajda, C. T.",
    title = "{Factorization, the light cone distribution amplitude of the B meson and the radiative decay $B \to \gamma \ell \nu_\ell$}",
    eprint = "hep-ph/0209216",
    archivePrefix = "arXiv",
    reportNumber = "CERN-TH-2002-228, SHEP-02-16",
    doi = "10.1016/S0550-3213(02)01066-0",
    journal = "Nucl. Phys. B",
    volume = "650",
    pages = "356--390",
    year = "2003"
}

@article{Beneke:2000wa,
    author = "Beneke, M. and Feldmann, T.",
    title = "{Symmetry breaking corrections to heavy to light $B$ meson form-factors at large recoil}",
    eprint = "hep-ph/0008255",
    archivePrefix = "arXiv",
    reportNumber = "PITHA-00-20",
    doi = "10.1016/S0550-3213(00)00585-X",
    journal = "Nucl. Phys. B",
    volume = "592",
    pages = "3--34",
    year = "2001"
}

@article{Braun:2003wx,
    author = "Braun, V. M. and Ivanov, D. Yu. and Korchemsky, G. P.",
    title = "{The $B$ meson distribution amplitude in QCD}",
    eprint = "hep-ph/0309330",
    archivePrefix = "arXiv",
    reportNumber = "LPT-ORSAY-03-63",
    doi = "10.1103/PhysRevD.69.034014",
    journal = "Phys. Rev. D",
    volume = "69",
    pages = "034014",
    year = "2004"
}

@article{Khodjamirian:2005ea,
    author = "Khodjamirian, Alexander and Mannel, Thomas and Offen, Nils",
    title = "{$B$-meson distribution amplitude from the $B \to \pi$ form-factor}",
    eprint = "hep-ph/0504091",
    archivePrefix = "arXiv",
    reportNumber = "SI-HEP-2005-01",
    doi = "10.1016/j.physletb.2005.06.021",
    journal = "Phys. Lett. B",
    volume = "620",
    pages = "52--60",
    year = "2005"
}

@article{Lee:2005gza,
    author = "Lee, Seung J. and Neubert, Matthias",
    title = "{Model-independent properties of the B-meson distribution amplitude}",
    eprint = "hep-ph/0509350",
    archivePrefix = "arXiv",
    reportNumber = "CLNS-05-1925",
    doi = "10.1103/PhysRevD.72.094028",
    journal = "Phys. Rev. D",
    volume = "72",
    pages = "094028",
    year = "2005"
}

@article{Beneke:2011nf,
    author = "Beneke, M. and Rohrwild, J.",
    title = "{B meson distribution amplitude from $B \to \gamma \ell \nu$}",
    eprint = "1110.3228",
    archivePrefix = "arXiv",
    primaryClass = "hep-ph",
    reportNumber = "TTK-11-51, SFB-CPP-11-56",
    doi = "10.1140/epjc/s10052-011-1818-8",
    journal = "Eur. Phys. J. C",
    volume = "71",
    pages = "1818",
    year = "2011"
}

@article{Feldmann:2022uok,
    author = {Feldmann, Thorsten and L{\"u}ghausen, Philip and van Dyk, Danny},
    title = "{Systematic parametrization of the leading B-meson light-cone distribution amplitude}",
    eprint = "2203.15679",
    archivePrefix = "arXiv",
    primaryClass = "hep-ph",
    reportNumber = "SI-HEP-2022-05, P3H-22-029, TUM-HEP-1388/22",
    doi = "10.1007/JHEP10(2022)162",
    journal = "JHEP",
    volume = "10",
    pages = "162",
    year = "2022"
}

@article{Feldmann:2014ika,
    author = "Feldmann, Thorsten and Lange, Bjorn O. and Wang, Yu-Ming",
    title = "{$B$-meson light-cone distribution amplitude: Perturbative constraints and asymptotic behavior in dual space}",
    eprint = "1404.1343",
    archivePrefix = "arXiv",
    primaryClass = "hep-ph",
    reportNumber = "SI-HEP-2013-18, QFET-2013-14, TTK-14-05, SFB-CPP-14-07, TUM-HEP-928-14",
    doi = "10.1103/PhysRevD.89.114001",
    journal = "Phys. Rev. D",
    volume = "89",
    number = "11",
    pages = "114001",
    year = "2014"
}

@article{Braun:2019wyx,
    author = "Braun, V. M. and Ji, Yao and Manashov, A. N.",
    title = "{Two-loop evolution equation for the $B$-meson distribution amplitude}",
    eprint = "1905.04498",
    archivePrefix = "arXiv",
    primaryClass = "hep-ph",
    reportNumber = "DESY-19-080",
    doi = "10.3204/PUBDB-2019-02451",
    journal = "Phys. Rev. D",
    volume = "100",
    number = "1",
    pages = "014023",
    year = "2019"
}

@article{Braun:2015pha,
    author = "Braun, V. M. and Manashov, A. N. and Offen, N.",
    title = "{Evolution equation for the higher-twist $B$-meson distribution amplitude}",
    eprint = "1507.03445",
    archivePrefix = "arXiv",
    primaryClass = "hep-ph",
    reportNumber = "DESY-15-108",
    doi = "10.1103/PhysRevD.92.074044",
    journal = "Phys. Rev. D",
    volume = "92",
    number = "7",
    pages = "074044",
    year = "2015"
}

@article{Descotes-Genon:2009jif,
    author = "Descotes-Genon, S. and Offen, N.",
    title = "{Three-particle contributions to the renormalisation of $B$-meson light-cone distribution amplitudes}",
    eprint = "0903.0790",
    archivePrefix = "arXiv",
    primaryClass = "hep-ph",
    reportNumber = "LPT-ORSAY-09-15",
    doi = "10.1088/1126-6708/2009/05/091",
    journal = "JHEP",
    volume = "05",
    pages = "091",
    year = "2009"
}

@article{Kawamura:2008vq,
    author = "Kawamura, Hiroyuki and Tanaka, Kazuhiro",
    title = "{Operator product expansion for $B$-meson distribution amplitude and dimension-5 HQET operators}",
    eprint = "0810.5628",
    archivePrefix = "arXiv",
    primaryClass = "hep-ph",
    doi = "10.1016/j.physletb.2009.02.028",
    journal = "Phys. Lett. B",
    volume = "673",
    pages = "201--207",
    year = "2009"
}

@article{Beneke:2021rjf,
    author = {Beneke, Martin and B{\"o}er, Philipp and Rigatos, Panagiotis and Vos, Kimberley Keri},
    title = "{QCD factorization of the four-lepton decay $B^-\rightarrow \ell \bar{\nu }_\ell \ell ^{(\prime )} \bar{\ell }^{(\prime )}$}",
    eprint = "2102.10060",
    archivePrefix = "arXiv",
    primaryClass = "hep-ph",
    reportNumber = "TUM-HEP-1317/21, Nikhef-2021-009",
    doi = "10.1140/epjc/s10052-021-09388-y",
    journal = "Eur. Phys. J. C",
    volume = "81",
    number = "7",
    pages = "638",
    year = "2021"
}

@article{Wang:2021yrr,
    author = "Wang, Chao and Wang, Yu-Ming and Wei, Yan-Bing",
    title = "{QCD factorization for the four-body leptonic B-meson decays}",
    eprint = "2111.11811",
    archivePrefix = "arXiv",
    primaryClass = "hep-ph",
    reportNumber = "TUM-HEP-1332/21",
    doi = "10.1007/JHEP02(2022)141",
    journal = "JHEP",
    volume = "02",
    pages = "141",
    year = "2022"
}

@article{Boeer:CKM2025,
    author = "P. B{\"o}er",
    title = "{Overview of recent progress in QCD factorisation for non-leptonic B-meson decays. Talk given at ``CKM 2025 -- 13th International Workshop on the CKM Unitarity Triangle''}",
    journal = "\url{https://indico.cern.ch/event/1440982/contributions/6583204/attachments/3134960/5562294/Boeer_CKM25.pdf}",
    year = "2025"
}

@article{BoeerNeubertStillger,
    author = "B{\"o}er, Philipp and Neubert, Matthias and Stillger, Michel",
    title = "{Factorization of Weak Annihilation Amplitudes in
Nonleptonic $B$-Meson Decays}",
    journal = "in preparation",
    year = ""
}

@article{Cornella:2026zkd,
    author = {Cornella, Claudia and Ferr{\'e}, Max and K{\"o}nig, Matthias and Neubert, Matthias},
    title = "{The simplest B decay, precisely}",
    eprint = "2601.14361",
    archivePrefix = "arXiv",
    primaryClass = "hep-ph",
    reportNumber = "CERN-TH-2026-006, MITP-25-075",
    doi = "10.1007/jhep06(2026)027",
    journal = "JHEP",
    volume = "06",
    pages = "027",
    year = "2026"
}

@article{Beneke:2020vnb,
    author = {Beneke, Martin and B{\"o}er, Philipp and Toelstede, Jan-Niklas and Vos, K. Keri},
    title = "{QED factorization of non-leptonic $B$ decays}",
    eprint = "2008.10615",
    archivePrefix = "arXiv",
    primaryClass = "hep-ph",
    reportNumber = "TUM-HEP-1277/20, MPP-2020-160",
    doi = "10.1007/JHEP11(2020)081",
    journal = "JHEP",
    volume = "11",
    pages = "081",
    year = "2020"
}

@article{Beneke:2019slt,
    author = "Beneke, Martin and Bobeth, Christoph and Szafron, Robert",
    title = "{Power-enhanced leading-logarithmic QED corrections to $B_q \to \mu^+\mu^-$}",
    eprint = "1908.07011",
    archivePrefix = "arXiv",
    primaryClass = "hep-ph",
    reportNumber = "TUM-HEP-1212/19",
    doi = "10.1007/JHEP10(2019)232",
    journal = "JHEP",
    volume = "10",
    pages = "232",
    year = "2019",
    note = "[Erratum: JHEP 11, 099 (2022)]"
}

\end{document}